# How do life, economy and other complex systems escape the heat death?

**Sorin Solomon \*, Natasa Golo**

Theoretical Physics Department, Racah Institute of Physics, Hebrew University, Givat Ram, Jerusalem 91904, Israel; sorin@huji.ac.il , natasa.golo@gmail.com .

**\*** Author to whom correspondence should be addressed; E-Mail co3giacs@gmail.com.
Tel.: +972 547 555 343.

**Abstract:** The primordial confrontation underlying the existence of our universe can be conceived as the battle between entropy and complexity. The law of ever-increasing entropy (Boltzmann H-theorem) evokes an irreversible, one-directional evolution (or rather *in*volution) going uniformly and monotonically from birth to death. Since the 19th century, this concept is one of the cornerstones and in the same time puzzles of statistical mechanics. On the other hand, there is the empirical experience where one witnesses the emergence, growth and diversification of new self-organized objects with ever-increasing complexity. When modeling them in terms of simple discrete elements one finds that the emergence of collective complex adaptive objects is a rather generic phenomenon governed by a new type of laws. These "emergence" laws, not connected directly with the fundamental laws of the physical reality, nor acting "in addition" to them but acting *through* them were called by Phil Anderson "More is Different", "das Maass" by Hegel etc. Even though the "emergence laws" act through the intermediary of the fundamental laws that govern the individual elementary agents, it turns out that different systems apparently governed by very different fundamental laws: gravity, chemistry, biology, economics, social psychology, end up often with similar emergence laws and outcomes. In particular the emergence of adaptive collective objects endows the system with a granular structure which in turn causes specific macroscopic cycles of intermittent fluctuations.

---

**1. The origin of universe space-time inhomogeneity**

The heat death concept was formulated by Rudolf Clausius in 1865 [1]. He stated that according to the second law of thermodynamics, any physical closed system tends toward the most probable equilibrium state: the state of maximum entropy. In the Wikipedia, the heat death is defined as *"… a suggested ultimate fate of the universe, in which the universe has diminished to a state of no thermodynamic free energy and therefore can no longer sustain processes that consume energy (including computation and life)"* [2].

Since its formulation, the heat death concept caused paradoxically an increasingly lively scientific effort to reconcile it with the empirically observed reality. It has been noted long ago that gravity has the capability of circumventing the heat death "theorem" because the gravity energy is not bounded from below. More precisely, a uniform mass distribution is not stable against small fluctuations. Zeldovich school *"explains the formation of the large-scale structure of the Universe as a consequence of the increase in the initially small matter density fluctuations due to gravitational instability.[…]this theory predicts that the growth rate of density perturbations D(t) is: $D(t) \sim \delta$ "* [3], i.e proportional to the (normalized) matter density fluctuations $\delta$ themselves.

Thus the smallest fluctuation that increases the density at some location will attract new mass to that location which in turn will attract new mass. The result will be an exponential increase of the microscopic fluctuations at certain singular locations. Thus the microscopic random events will develop towards macroscopically relevant effects: vast regions of relatively empty space punctuated by very singular locations with enormously large density: galaxies, black holes, etc.

This scenario is very typical for the other applications in this paper: a quantity whose growth rate is proportional to itself is bound to undergo a very singular spatio-temporal localized dynamics.

A very wide range of systems: viruses, animals, nations, parties, cultures, prices, follow similar spatio-temporal singular patterns due to their similar stochastic autocatalytic behavior:

$$\frac{dW(x,t)}{dt} = \eta(x,t) \cdot W(x,t) \tag{1}$$

where $\eta(x,t)$ is a random multiplicative factor.

Thus the scenario resulting from Zeldovich's approximation is only the first in a series of phenomena that give rise to structure, dynamics and life instead of a uniform state converging monotonically towards a stationary heat death end.

## 2. Survival implications of the inhomogenous growth

Let us start with a trivial example (the list of references includes enough technical caveats that we wish to avoid at the beginning of this exposition). Imagine that one starts with an arbitrarily large universe with a constant density of living / civilized agents. And then imagine that averaging over the various locations one finds that their density decays at an average rate of 50% per billion years.

Would life disappear from such a universe? *Not necessarily!*

Suppose the universe is spatially inhomogeneous:

- in 99% of its volume the conditions are such that the density shrinks at a rate of 48% per billion years. The fraction of the population in the shrinking region evolves thus according to the equation:

$$\frac{dW(99, time)}{dt} = 0.48 \cdot W(99, time). \tag{2}$$

So,

$$W(99, time) = 0.99 \cdot 0.48^{time} \cdot W_o \tag{3}$$

where *time* is measured in billion years and $W_0$ represents the initial total population

- in the rest of 1% of the locations increases at a rate of 200%. The fraction of the population in the growth region evolves thus according to the equation:

$$\frac{dW(1, time)}{dt} = 2.0 \cdot W(1, time). \tag{4}$$

So,

$$W(1, time) = 0.01 \cdot 2^{time} \cdot W_o \tag{5}$$

i.e. doubling every billion years.

What is the actual fate of life in such a universe?

$$W(total, time) = 0.98 \cdot W(99, time) + 0.01 \cdot W(1, time) = 0.99 \cdot 0.48^{time} \cdot W_0 + 0.01 \cdot 2^{time} \cdot W_0 =$$
$$= (0.99 \cdot 0.48^{time} + 0.01 \cdot 2^{time}) \cdot W_0 \tag{6}$$

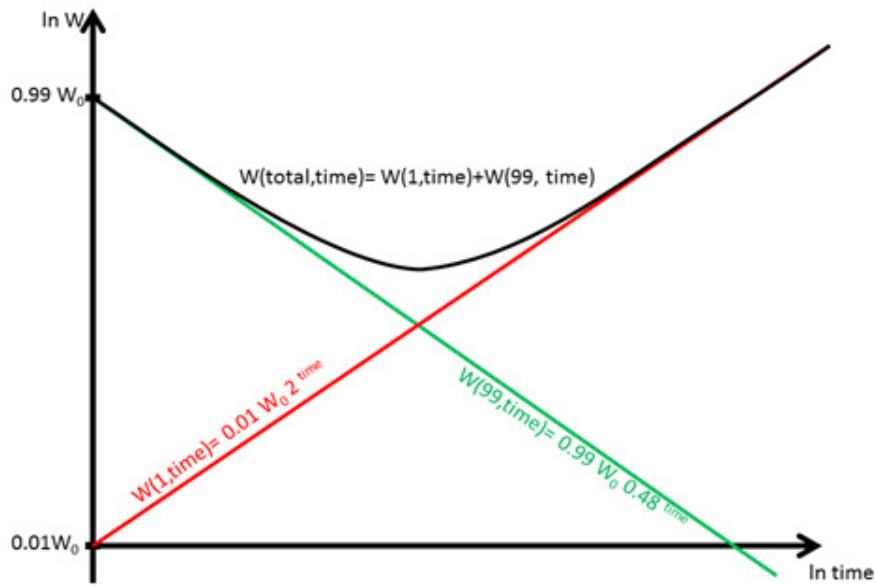

**Figure 1.** Illustrates the evolution of the population in the universe as expressed by the dynamics Eq. 2-6. The main features are the "**crossing exponential**": at the beginning of the process, 99% of the population resides in the negative growth regions and 1% in the growth regions. Thus the *dynamics* of the system as a whole is initially dominated by the **decaying exponential**. As time passes the initially

small 1% population grows exponentially and the relative sizes of the 2 populations **cross**: one on its way to extinction and the other on its way to expansion. Eventually the population of the 1% growth volume constitutes nearly the entire total population. At this stage the dynamics of the system reduces entirely to a **growing exponential**.

According to Eq. 6 and as visualized in Fig. 1, at the beginning the average density will decay exponentially as $0.99\, W_0\, 0.48^{time}$ (where time is measured in billion years) because the contribution $0.01\, W_0\, 2^{time}$ of the growing 1% is negligible.

However after the first 7 billion years, the decaying 99% of the universe would have been reduced roughly by a factor of $1/128 \sim (0.48)^7$ and its contribution

$$0.99 \cdot 0.48^7 < 0.01$$

will be dominated by the population of the growing 1% which by now increased by a factor of $128 \sim 2^7$ to

$$0.01 \cdot 128 > 1 >> 0.99 \cdot 0.48^7$$

Thus after a very small number (4-7) of cycles, the civilizations will be concentrated in the 1% growth volume and the average density will grow at a rate of 200% (i.e. doubling) per billion years (as opposed to the average growth rate over the locations which remains 0.5).

Actually, life is even more resilient than that. It survives even if the 1% growth volume is shrinking. For instance, if one assumes now that every 2 billion years the populations of half of the currently growth locations are destroyed (or self-destroy by endogenous effects), the living / intelligent agents will still survive. The growth rate of the average density will still exceed 1. Indeed, in the 2 billion years during which ½ of the populations disappear, the populations at the surviving 1/2 of the growth locations will increase by a factor of $2^2 = 4$. Thus, all in all, one obtains a growth factor of the universe average

$$½ \cdot 0 + ½ \cdot 4 = 2 \text{ each 2 billion years.}$$

(while the average growth rate over all universe locations decreases from the initial 0.5 towards 0.48).

This is the kind of singular dynamics that defeats the naïve conclusion that there is no other way for the universe but to converge towards an equilibrium in which nothing happens macroscopically and reach a thermodynamic uniform state/ heat death.

While the cosmological implications are far away from daily life and somewhat hypothetical we will show that the same mechanisms are of high relevance for the survival of animal populations and economic systems in real life. The "crossing exponentials" effect: the decaying of the old dominant components of the system and the emergence of the new from a tiny initially negligible seed is so intrinsic to life that it transcends our intellectual interest and it moves us instinctively.

## 3. Examples of autocatalytic dynamics in various fields

Eq. 1 and the example in Section 2 (Eq. 6, Fig. 1) represent/stand for a much wider mechanism appearing under a multitude of names in different domains: chemistry (under the name of auto-catalysis), biology (reproduction / multiplication, species perpetuation), economics (profit, returns, rate of growth, economy of scales). In fact we will argue that this is the dominant mechanism which insures the generic emergence of macroscopic collective effects in systems that include many interacting elements. In the present text we will use the term **auto-catalysis** which was used with similar meaning by [4] in describing the emergence of life from generic chemistries.

As opposed to the usual stochastic systems in which the basic dynamics changes typically the individual microscopic quantities by additive steps (e.g. a molecule receiving or releasing a quanta of energy), the auto-catalytic microscopic dynamics involve multiplicative changes (e.g. the market worth of a company changes by a factor (gain) after each elementary transaction, virus populations multiply (or die) at a given rate, etc). At system level, autocatalytic loops can appear when a change in the state of an agent causes the same change in other individuals (contagion). Particularly interesting forms of autocatalyticity are the instances where the effects are enhancing their own causes. Such logically circular structures have in principle the potential to generate novelty, creativity and free will by initiating self-referential infinite loops (e.g. systems reacting on their own components) that transcend the initial logical set-up [5, 6, 7].

Examples of autocatalytic dynamics can be found in each scientific and human domain at the basis of most of the complex systems:

- in philosophy, the self-fulfilling prophecy, a prediction influencing the behavior of humans to the effect of the prediction fulfillment, was described by Carl Popper, who called it the Oedipus effect [8] :

    *"By the name 'Oedipus effect,' I wish to allude to a certain aspect of the story of Oedipus whose fate was predicted by the oracle, with the result that the prediction induced Oedipus' father to those very actions which ultimately brought about the predicted events. The effect of a prediction (or a similar piece of information) upon the events or objects to which the prediction refers-for example, by promoting or by preventing the predicted events - I have called the 'Oedipus effect '. One would expect the Oedipus effect to be a peculiarity of the social sciences and their subject matter, and that it is completely foreign to physics; but this is not the case."*

    Popper finds roots of the Oedipus effect in the principle of interdeterminacy: an individual's self-information of its own state at $t$ , in order to be correct and up to date, would have to contain, as one of its parts, a physical description of that very self-information, using physical symbols of the same kind as in the rest of its information. But this is impossible for any finite piece of information; and an infinite piece of information cannot, of course, be completed at any instant of time, so that it cannot become the basis of a prediction.

Popper's position has had the support of Niels Bohr [9]:

*"In fact, the use which we make of words like "thought" and "feeling," or "instinct" and "reason" to describe psychic experiences of different types, shows the existence of characteristic relationships of complementarity conditioned by the peculiarity of introspection. Above all, just the impossibility in introspection of sharply distinguishing between subject and object as is essential to the ideal of causality would seem to provide the natural play for the feeling of free will."*

This makes a feedback loop available for the psyche: the observer influences the observed which is itself.

- In bibliometrics/epistemology, the collective dynamics in knowledge networks, i.e. the propagation of ideas among the scientists has been discussed in the citation networks growth model of [10]. This ostensibly bibliometric work is actually expressing in dry numbers the autocatalytic effects governing the spread of ideas and knowledge within the collective intelligence [11] of entire human communities. Here too the emergence of knowledge is the result of collective effects governed by singular centers of autocatalytic growth [12].
- In ecology, one has advanced the *Gaia Hypothesis*: Gaia is the name of the biosphere regarded as a living "superorganism" [13]. It postulates that the whole of life, air, ocean, and rocks evolves together as the earth biosphere feedback enhances the growth of its individual components (flowers, trees, animals). A mathematical metaphor for the Gaia theory is the Daisyworld model [13]. The Daisyworld model postulates feedback loops between living species and the environmental conditions. One assumes there are 2 kinds of daisies: black and white (in some versions also all shades of gray). The black daisies like to live in the cold climate, because being black they absorb more energy from the sun. Thus if one starts with a cold climate, the black daisies will take over the white ones. At the planetary level having the earth surface covered by black, will retain in the earth system an increasing quantity of the solar energy reaching the earth. This will result in an increasing temperature of the earth. In the resulting warmer climate the white daisies which keep out of overheating by reflecting the sun light will be advantaged. As the white daisies start to predominate, at the planetary level, the earth will reflect part of the solar energy reaching it. The climate will cool down again and the black daisies come to flourish again. This feedback loop between the living individuals and the earth (Gaia) system as a whole are postulated by the Gaia hypothesis to bring the Gaia system as well as its individual components to an optimal state. The mechanism can be reinterpreted conceptually as an adaptive self-sustaining property of the Gaia system. The basic idea was further improved by adding many layers of complexity and used to predict the impact that the humans can have on the biosphere.
- The biologist Robert May postulated that the dynamics of populations, wealth, politics is governed by the logistic equation [14]:

*"I would therefore urge that people be introduced to [the logistic equation] early in their mathematical education. … Not only in research, but also in the everyday world of politics and economics"*

The crucial (linear) term of the logistic equation has a form identical with Eq. 1. It was shown in the context of the AB model described below [15] that this term has the capacity to amplify the microscopic local fluctuation as to generate spontaneously macroscopic adaptive collective objects:
- o  proto-cells that intermediate the appearance of life.
- o  Species that insure life's resilience
- o  Herding effects by which the populations select the optimal environment (rather than vice-versa as prescribed by the usual Darwin natural selection).
- in social psychology "Deutsch's Crude Law of Social Relations" [16] states that:

*"The characteristic processes and effects elicited by a given type of social relationship (e.g., cooperative or competitive) tend also to elicit that type of social relationship; and a typical effect of any type of relationship tends to induce the other typical effects of the relationship".*

In [17] a model have been developed which, generalizing the AB model, describes the emergent consequences of the assumptions of the Crude Law, as applied to interactions within and between social groups. The results of the simulations show collective phenomena that could not be anticipated in the original formulation of the law that was limited only to individual psychological facts: Autocatalytic herding and spatial localization mechanisms are found to be crucial for determining global dynamics of conflict. Thus the Crude Law implies that conflicts grow exponentially in places with the highest incompatibility of interests, and spread from these places. Conflicts escalate to intractability by altering social orientations in the areas of highest intensity. Disruption of local constraints (e.g. movement through free travel) can paradoxically reduce the conflict. Simulations also suggest that seemingly intractable conflicts can be resolved by creating autocatalytic positive processes in the vicinity of the conflict center.

The last examples have been formalized and solved in the context of the AB model [15] which we describe below.

## 4. The AB model

An agent based model that describes very economically a generic road to escaping the heat death sentence is the AB model. The model consists of 2 types of agents on a lattice:
- A's which only may diffuse by jumping randomly between neighboring locations , and
- B's which in addition to diffusing can multiply (generate new B's) when sharing the same location with an A.

At the first sight, the model corresponds at the macroscopic level to the logistic equation and leads to a death-like uneventful spatially uniform state. However, the effects of microscopic granular structure of the A agents are not lost at the macroscopic level. On the contrary, instead of an uniform macroscopic state one predicts the spontaneous emergence of macroscopic adaptive collective B objects which defeat the entropy and survive indefinitely by effectively exploiting the fluctuations implicit in the microscopic granularity of the A's. Beyond its importance of principle, this simple model explained and predicted accurately empirical phenomena in microbiology, population dynamics, economics, finance, real estate, social psychology etc.

The *AB* Model is a discrete stochastic spatial extension of the logistic system: the A's are the catalysts that make the conditions for the reactants *B's* to grow. Both types of agents (*A* and *B*) come in discrete quantities (0,1,2,3,..) and can diffuse with their own diffusion rate equal to $D_a$ and $D_b$, respectively. The agents of type *A* cannot multiply nor die, while the *B* agents, can die but can also multiply in the presence of *A*'s. In chemical reaction-diffusion notations the system is:

$$A + B \xrightarrow{\lambda} A + B + B \qquad \lambda \text{ proliferation rate (autocatalyticity)} \qquad (7)$$

$$B \xrightarrow{\mu} 0 \qquad \mu \text{ decay (death rate)} \qquad (8)$$

$$B + B \xrightarrow{\chi} B \qquad \chi \text{ inhibition (not used in most of the paper)} \qquad (9)$$

If one looks at the local densities $a(x,t)$, $b(x,t)$ of the A and B agents as continuous functions governed by the differential equation implied by the reactions 7-9 one obtains the logistic equation proposed by Malthus and Verhulst [18, 19] 200 years ago:

$$\frac{db(x,t)}{dt} = \eta(x,t) \cdot b(x,t) - \chi b^2(x,t) + D_b \Delta b(x,t) \qquad (10)$$

where $\eta(x,t) = \lambda a - \mu$ is in the continuous approximation constant because the A's diffusion

$$\frac{da(x,t)}{dt} = D_a \Delta a(x,t) \qquad (11)$$

causes the A density to converge very fast to a constant *a* in space and time. The constant *a* is fixed by the initial conditions because the A agents are not created or destroyed in time.

For simplicity, as in the previous cases, we will concentrate on the effects of the first (Malthus) linear term in Eq. 10 i.e. neglect below the reaction 9.

The naive continuous, space-time uniform model 10-11 gives a very simple behavior, wherein depending on the average death rates and birth rates, the density of B would either have a spatially uniform growing exponential trend (for $\lambda a > \mu$) or uniformly decaying exponential (if the $\lambda a < \mu$):

$$\frac{db}{dt} = (\lambda a - \mu)b > 0 \rightarrow population\ growth$$

$$(12)$$

$$\frac{db}{dt} = (\lambda a - \mu)b < 0 \rightarrow population\ decay$$

However, the discrete AB model result is quite different. Instead of the monotonous exponential, it displays even for $\lambda a < \mu$ a quite lively and thriving B population. The *B's* in the 7-9 model self-organize spontaneously in herds / island-like spatial structures. This is related with the dictum by Phil Anderson [20]:

*"Real world is controlled … by the **exceptional**, not the mean; by the **catastrophe**, not the steady drip".*

The point is that the microscopic fluctuations in the (macroscopically uniform) distribution of the A's lead the spontaneous emergence and growth of macroscopic adaptive herds / islands of B.

Clearly, the growth rate of the B population $b(x,t)$ will be the largest at the locations with the largest number $a(x,t)$ of A's. More precisely neglecting for a moment the motion of the A's (i.e. the time dependence of $a(x,t)$), the evolution of the B's will be given by a sum of exponentials:

$$b(t) = b(0) \sum P(a(x,t)) e^{(a(x,t)\lambda - \mu)t} \qquad (13)$$

The places with the largest values of $a(x,t)$ are the rarest, their probability $P(x,t)$ tends to 0 (cf. Eq. 17) when $a(x,t)$ tends to infinity:

$$P(a(x,t)) \xrightarrow{a(x,t) \to \infty} 0 \qquad (14)$$

But they have the fastest growth rate:

$$e^{(a(x,t)\lambda - \mu)t} \xrightarrow{a(x,t) \to \infty} \infty \qquad (15)$$

The dynamics of the AB systems is thus dominated by the rarest and most singular events. There is no wonder that the differential equations Eqs. 10-11 that assume a smooth behavior of the densities $a(x,t)$ and $b(x,t)$ are missing the main point of the logistic dynamics: the emergence of singular growth centers at the largest A fluctuations.

As the A's are diffusing, the locations with largest $a(x,t)$ 's are moving, disappearing and appearing and so do the terms in the sum Eq. 13. The non-trivial effects proved in [15, 21] is that the B herds / islands are capable to discover, follow, adapt their position and exploit the fortuitous A fluctuations to such a degree that in 1 and 2 dimensions the B population never dies. This phenomenon has been studied in great detail in various regions of the parameter space and with focus on various mechanisms as described in the following sub-sections.

*4.1. Renormalization group analysis*

The Renormalization Group analysis expresses the fact that the correlations between the locations of A and B renormalize the effective birth rate $\lambda(s)$ to increasing values as one increases the size $s$ of the system [22].

In simple terms, if the average densities of A and B are respectively $a$ and $b$, then Eq 10 assumes naïvely that the probability for an A and B to meet is $a \cdot b$ and that consequently the probability per unit time for a generating a new B is $\lambda a \cdot b$. This is actually false: the B particles tend to be born in the vicinity of the A's. Thus, the probability for an A and B to be found on the same site and produce a new B far exceeds $\lambda a \cdot b$. Once that B is produced on the same site with the A, the probability of producing yet another B further increases. Of course, this argument discriminating between the local density of B's in the A neighborhood and between the average B density does not hold for a degenerate 1x1 lattice and is not very powerful for a 2x2 lattice either. However, as one considers larger and larger lattices, the correlations between the A and B fluctuations above their averages increase and affect more and more the naïve proliferation rate $\lambda a \cdot b$.

This can be taken formally into account by defining a "renormalized" value of the B reproduction term $\lambda(s) a \cdot b$ which depends on the scale of the system $s$. Thus as one increases the lattice of dimensions $sxs$ the effective constants $\lambda(s)$ and $(\lambda a - \mu)(s)$ will change their values as seen in the inset of Fig. 2. This motion of the effective constants as a function of the scale $s$ is called renormalization flow. As seen in the figure, for large enough 2 dimensional systems the effective value $(\lambda a - \mu)(s)$ becomes positive. Thus, the spatial correlations between A and B large density regions ensure that irrespective of the averages $a, b$, of the original ("bare") constant $\lambda$, and the other parameters of the system $\mu$, $D_a$, $D_b$ the B population always increases in a large enough system: "life always wins on a surface (and on a line)". In higher dimensions, B population death is possible but in much more restrictive regions than implied by Eq. 12.

Fig. 2 substantiates the claims above: the same parameters $\mu$, $\lambda$, $a$, $D_a$, $D_b$ which lead in small size (30x30) systems to B extinction, lead for large enough system size (100 x 100) to B survival and thriving.

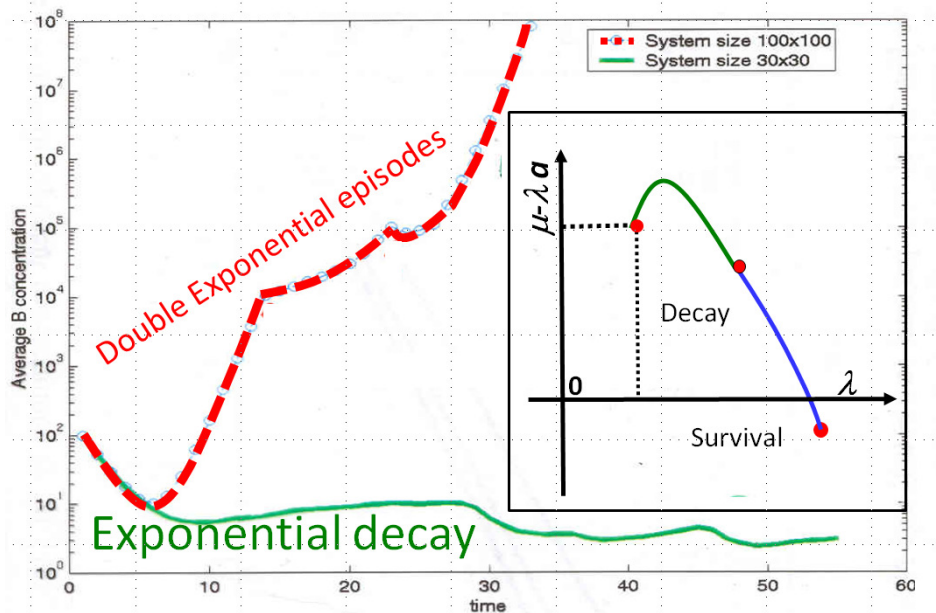

**Figure 2 The inset** represents the renormalization flow for a system defined locally by the "bare" constants $\lambda$, $a$ and $\mu$. The actual independent combinations considered are $\lambda$ represented on the horizontal axis and $\mu - \lambda a$ represented on the y axis. The initial values for the minimal system size (1x1) are represented by the leftmost dot. They correspond to $\mu - \lambda a > 0$ i.e to B disappearance according to the naïve Eq. 12. The coordinates of the second dot represent the effective values of the $\lambda$ and $\mu - \lambda a$ for a 30x30 system. The line connecting between the first and second circle represents the renormalization flow of the effective parameters as one increases the size of the system from 1x1 to 30x30. The third dot represents the effective parameters for a 100x100 lattice. One sees that while for 30x30 the renormalization group still predicts B disappearance, $\mu - \lambda a > 0$, for 100x100 the theoretical prediction is that life wins, $\mu - \lambda a < 0$.

**The main** figure represents the time evolutions of the actual numerical simulations of the AB system. One sees that indeed, as predicted by the theory, the B population decays for the 30x30 system (solid line) but survives and grows for the 100x 100 system (dashed line). Note however the fine structure of the dashed line: "crossing exponentials" separated by cusp discontinuities discussed in the next sections.

*4.2. The random branching walks analysis [21, 23]*

The survival of the B population can be connected to the Polya constant $P_d$. For a given dimension $d$, the Polya constant $P_d$ is the probability for a randomly walking A to return to the origin of the walk if allowed enough time. It has been proved that the condition for the B survival is:

$$\lambda / D_a > 1 - P_d \tag{16}$$

where $D_a$ parametrizes the diffusion of the A's: it is the probability per unit time for an A to jump on a neighboring site.

The relation 16 quantifies the fact that the Polya constant insures that the large fluctuations in $a(x,t)$ endure over long enough time periods as to allow the B population to follow them and thereby survive.

We choose here to describe in more details the proof based on [21] branching random walks analysis. This is because it allows quantitative evaluation [23] of the conditions and frequency of the emergence of singular growth centers. The proof uses Polya theorem which states that the random walks in 2 and less dimensions are recurrent. By recurrent, one means that the random walker stays in a finite neighborhood of its site of origin and in fact it returns an infinite number of times to its original site. To understand the significance of this property, note that in 3 and higher dimensions, the random walks are not recurrent: the random walker goes to infinity and is not anymore returning to the neighborhood of the origin.

The proof is formally about the expected number of B's at a specific site, but in practice applies to any site in its neighborhood. Thus it is the formal expression of the fact that the changes in the A configurations are intrinsically slow enough to allow the birth of new B's to ensure that the B herds succeed to follow the changes in the A configuration.

Let us define $P_d(t)$ the probability (in $d$ dimensions) that a random walker returned at least once to its site of origin by time $t$. With this definition the Polya constant is : the probability $P_d(\infty)$ that an A eventually returns to its site of origin. In one dimension d=1 the asymptotic behavior is $P_1(t) \approx 1 - 1/\sqrt{t}$ and in 2 dimensions $P_2(t) \approx 1 - 1/\ln(t)$, therefore $P_1(\infty) = P_2(\infty) = 1$. As mentioned above, for $d > 2$, $P_d(\infty) < 1$. For example, the value of the Polya constant in 3 dimensions is $P_3(\infty) = 0.3405373$.

Kesten and Sidoravicius [21] proved that on large enough 2 dimensional surfaces, regardless of the choice of the parameters $\lambda$, $\mu$, $D_b$, $D_a$, $a$, the total B population $b(t)$ always grows. The proof below is based on their analysis, though we ignore their sophisticated discussion of the existence of the process etc, and we only concentrate on the aspects that describe the mechanism involved. Moreover, unlike the mathematicians, which were content with the finding that the total growth in 1 and 2 dimensions is even faster than exponential, we try to quantify further the conditions for growth:

- In $d < 2$ we give an evaluation of the minimal size V of the system necessary to avoid sure death of the B population.
- In $d > 2$ dimensions we give the condition for the B population survival (for large enough systems) as a lower bound for the ratio between the B proliferation rate and the A diffusion rate: $\lambda / D_a > 1 - P_d$.

The key point which makes the proof possible is to realize that the motions of the different A's are independent of one another. Moreover they are independent on the birth, death and motion of the B's. This will allow us below to factorize as independent factors their respective multiplicative contributions to the expected number of B's on a site $\mathbf{b}(x,t)$. Here and in the sequel we denote by a bold letter $\mathbf{b}$ the expected value of a quantity $b$.

Thus we will be able to treat different contributions to the growth of $\mathbf{b}(x,t)$ at a given site separately, by taking advantage of the fact that motion of each of the **A**'s is independent of:

- the B's life and death,
- the B's migration,
- the other A's presence,
- the immigration of B's.

To exploit this we limit the analysis along the following lines:

- We assume each A and each B has a uniform probability to be initially, at time $t=0$ on any of the sites $x$ of the $d$ - dimensional lattice / network. In particular this means that $a(x,0)$ has a Poisson probability distribution:

$$P(a(x,0)) = a^{a(x,0)} e^{-a} / a(x,0)! \qquad (17)$$

where $a$ is the average of the number of A per site in the system.
- We concentrate on the growth of the expected number $\mathbf{b}(x,t)$ of B agents at a given site $x$.
- We consider only the contribution to $\mathbf{b}(x,t)$ by the $a(x,0)$ agents A that resided at $x$ at the beginning of the process $t = 0$. The contributions by A's that were at $t = 0$ on different sites $y \neq x$ and arrive by chance at some stage $t \neq 0$ at $x$ are neglected.

- We neglect the B's that diffused away from $x$ and are returning during the following steps. I.e. the diffusion contributes just a decay factor $e^{-D_b t}$ to $\mathbf{b}(x,t)$. Thus, it can be lumped with the B death to give a decaying exponential contribution $e^{-(\mu+D_b)t}$.
- Note that all the neglected contributions above are increasing $\mathbf{b}(x,t)$ so by enforcing them we will obtain a lower bound for $\mathbf{b}(x,t)$ thus allowing us to prove exponential increase.
- The contributions of the original A's at $x$ are independent and therefore contribute a factor: $e^{a(x,0)E(t)}$, where $e^{E(t)}$ is the expected factor by which one single A contributes to $\mathbf{b}(x,t)$.

With all these conventions one gets a lower bound to $\mathbf{b}(x,t)$:

$$\mathbf{b}(x,t) > e^{a(x,0)E(t)-(\mu+D_b)t} \tag{18}$$

Thus, if we will prove that

$$E(t) > E \cdot t \tag{19}$$

for some positive constant $E$, then all the sites that respect

$$a(x,0) > (\mu + D_b)/E \tag{20}$$

will have an exponentially growing $\mathbf{b}(x,t)$ expectation. According Eq. 17 this means that at least a fraction:

$$P(a(x,0) = (\mu + D_b)/E) = a^{(\mu+D_b)/E} e^{-a} /[(\mu + D_b)/E]! \tag{21}$$

of the sites of the system will increase exponentially. Thus, even if we neglect all the other sites of the system, just this subset ensures that for $t \to \infty$ the sum / average $b(t)$ of all the $b(x,t)$'s increases at least exponentially (with the price of a very small fixed factor Eq. 21).

Of course if the volume (number of sites in the system) is less then

$$V < 1/P(a = (\mu + D_b)/E) = a^{-(\mu+D_b)/E} e^{a} [(\mu + D_b)/E]! \tag{22}$$

the expected number of growing sites will be less than 1 and the B population will disappear very probably. So by obtaining the value of $E$ we will be able not only to prove the survival and growth of B's in infinite size systems but also to estimate what is the sufficient size of the system V Eq. 22 in order to obtain B population growth given a set of parameters $a$, $\mu$, $D_b$, $D_a$, $\lambda$.

Thus we are left now with the crucial question: does or does not each of the $a(x,t=0)$ agents A that at time $t=0$ resided at site $x$ make an expected exponential contribution of at least $e^{Et}$ (with some constant $E$) to the expected growth of $\mathbf{b}(x,t)$?

So from now on we will focus only on the effect of a single A to the growth of B's on this particular location. The A's contribution to the growth of $\mathbf{b}(x,t)$ is dependent of 2 factors: the probability that an A will leave the site, the probability of its return to that site, and the duration of stay of each of the 'visits'.

The duration of the visits (including the duration of the original stay at the origin before the first departure) is distributed by the probability distribution

$$P_{Stay}(t)dt = D_a e^{-D_a t} dt \tag{23}$$

We distinguish now 2 cases:

1) If the proliferation rate $\lambda$ is large enough, $\lambda > D_a$, then we already have an expected time exponential factor contribution to $\mathbf{b}(x,t)$:

$$e^{E(\tau)} = \int_0^\tau P_{Stay}(t) e^{\lambda t} dt = \int_0^\tau D_a e^{(\lambda - D_a)t} dt = \frac{e^{(\lambda - D_a)\tau} - 1}{(\lambda/D_a) - 1} \propto e^{(\lambda - D_a)\tau} \tag{24}$$

i.e. an exponential increase with $E(t) = E \cdot t = (\lambda - D_a)t$ as required by Eq. 19.

2) In the case where $\lambda < D_a$, the integral does not grow exponentially and should rather be written as

$$e^{E(\tau)} = \int_0^\tau D_a e^{(\lambda - D_a)t} dt = \frac{1 - e^{-(D_a - \lambda)\tau}}{1 - (\lambda/D_a)} < \infty \tag{25}$$

Thus, considering only the possibility of A staying arbitrarily long time at its site of origin is not enough. Rather one should consider the recurrence of such stays by multiplying Eq. 25 with the probability of return $P_d(t)$ after a certain time $T$.

The expected contribution to $\mathbf{b}(x,t)$ of $N$ such events is their probability $(P_d(t))^N$ times their contribution (Eq. 25 multiplied with itself N times):

$$e^{E(t)} = \left[ P_d(T) \frac{1 - e^{-(D_a - \lambda)\tau}}{1 - \lambda/D_a} \right]^N, \tag{26}$$

where we denoted the total time that such a process takes by $t = N(T + \tau)$,

$$e^{E(t)} = \left[ P_d(T) \frac{1-e^{-(D_a-\lambda)\tau}}{1-\lambda/D_a} \right]^{t/(T+\tau)} = \left\{ \left[ P_d(T) \frac{1-e^{-(D_a-\lambda)\tau}}{1-\lambda/D_a} \right]^{1/(T+\tau)} \right\}^t \qquad (27)$$

Now it is left to see that for an appropriate choice of $T$ and $\tau$, the term in the curl brackets can be larger than 1. This is not-so-difficult since $P_d(T)$ and $1-e^{-(D_a-\lambda)\tau}$ can be arbitrarily close to 1 for large enough $T$ and $\tau$ while $\frac{1}{1-\lambda/D_a}$ is larger than 1. Raising it to the positive $1/(T+\tau)$ power does not change this property. So the curl bracket can be written as $e^E$ with $E$ a positive constant. So the condition 19 is fulfilled:

$$e^{E(t)} > e^{Et}, \qquad (28)$$

which finishes the proof. In particular one can find that with judicious choice of $T$ and $\tau$, one can have

$$E \sim \frac{\lambda}{D_a} e^{-2D_a/\lambda} \qquad (29)$$

which through Eq. 22 gives an estimation of the system size $V$ necessary to realize $\mathbf{b}(t)$ growth in a given sample.

*4.3. The directed percolation*

In another field theory application [16] the AB model has been mapped on the directed percolation problem. The survival of life in this context is related to the fact that as the A fluctuations move / disappear, the B herds / islands around them are capable to survive enough to follow their motion or until they meet a new emergent A fluctuation [23]. Thus, in spite of each of the growth centers having a finite life-time, they succeed to maintain an unbroken chain of survival related to the directed percolation phase transition [23]. With very simple mechanical assumptions one obtained a rigorous mathematical metaphor for the eternal transmission of life between the generations of otherwise mortal individuals.

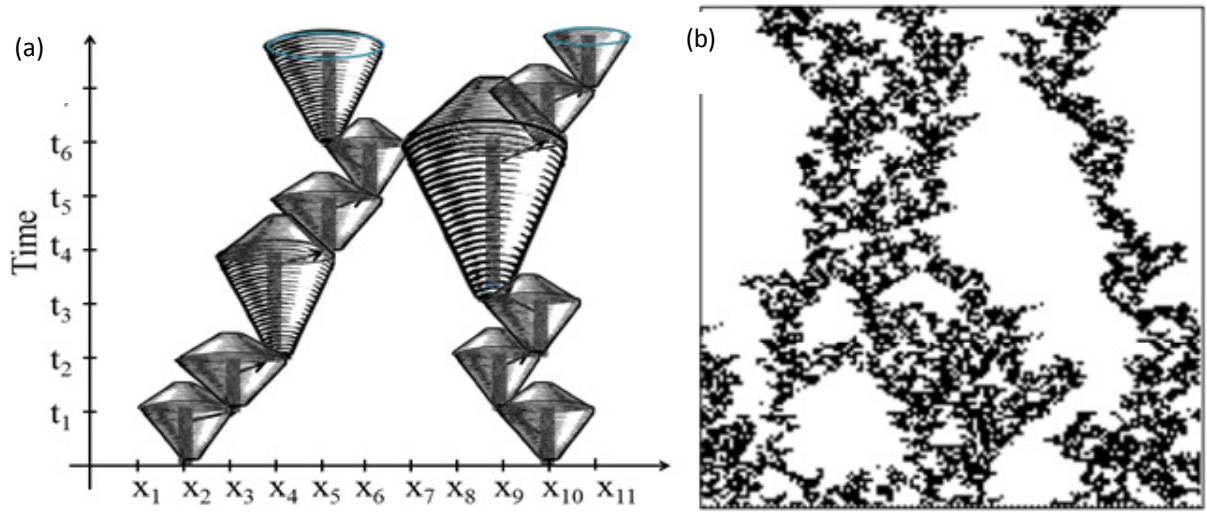

**Figure 3. (a)** The dynamics of the A fluctuations and their associated B herds can be connected to the formalism of directed percolation. The vertical direction denotes time and the horizontal one space. The black vertical cylinders represent regions in space which for a time have a large positive A fluctuation that can sustain a growing "island". The herd / island grows at constant speed until the A fluctuation that sustained it disappears. Then the herd / island starts shrinking. If by the time the old herd disappeared there appears another A fluctuation in the neighborhood that can help multiply some of the surviving B's from the old herd, then a new island / herd is generated and starts expanding. This kind of dynamics can be abstracted to a percolation system where contagion (B transfer from the old to the new A fluctuation) can take place only in the "positive" (causal, time increasing direction).
**(b)** (from [24]) depicts such a "directed percolation" system. According to the directed percolation theory one needs a particular density of susceptible sites (A fluctuations) to support the process such that contagion of new sites never stops. In the AB model in $d < 2$ dimensions the fact that the A fluctuations change very slowly (A's remain in their finite neighborhood of origin rather than getting lost to infinity) insures that the disappearing A fluctuations have nearby descendants. Thus while each of the collective objects have a finite life time, life itself finds always new collective objects to embody itself in.

## 5. Properties, predictions and applications of the AB model

The AB model has a very rich set of predictions which are realized in many empirical systems. However they are sometimes difficult to understand by the practitioners in the target field of application because of the intricate technicalities of the renormalization group, branching random walk and Monte Carlo simulation techniques. Therefore in this chapter we will explain the main predictions and their empirical applications in the context of a very simple and visually intuitive approximation.

*5.1 The single A approximation [25]*

As was shown in [15] the B's self-organize in herds / islands-like structures that follow and exploit the positive A fluctuations. These collective objects have strong adaptive properties and are much more efficient in exploiting the available resources than each of the individual B's composing them.

The field theory computations, random walk theorems and the computer simulations give each a different aspect of the intuitions behind the rich behavior of the AB model. In this section we introduce a new set of intuitions with yet a simpler way to look at the AB model: the single A approximation.

Suppose one has a very low density $a$ of A's:

$$a << \mu/\lambda < 1 \tag{30}$$

In fact suppose the A's are so sparse that they seldom meet on the same site (and we neglect those events).

- The first inequality in Eq. 30, implies that according to the continuum approximation Eq. 10-12 the total B population $\mathbf{b}(t)$ will decay exponentially as

$$\mathbf{b}(t) \sim e^{-(\mu-a\lambda)t} \tag{31}$$

- On the other hand the second inequality in Eq. 30, $\mu < \lambda$ implies that even one A at a location $x$ is enough to insure the exponential growth of the expected B population at that $x$ as:

$$\mathbf{b}(x,t) \sim e^{(\lambda-\mu)t} \tag{32}$$

The B population expectation $\mathbf{b}(y,t)$ at the sites y where there is no A agent will decay exponentially as

$$\mathbf{b}(y,t) \sim e^{-\mu t} \tag{33}$$

For small enough times when one can use the approximation $e^x \cong 1 + x$ one obtains for the discrete system the same result as in the continuum approximation (Eq. 31):

$$\mathbf{b}(t) \cong a\mathbf{b}(x,t) + (1-a)\mathbf{b}(y,t) \cong a[1+(\lambda-\mu)t] + (1-a)[1-\mu t] \cong 1+(a\lambda-\mu)t \cong e^{-(\mu-a\lambda)t} \tag{34}$$

i.e. an *exponential decrease*. However, for large times the term 33 becomes negligible and the dynamics is dominated by the sites where the A reside Eq 32:

$$\mathbf{b}(t) \cong a\,\mathbf{b}(x,t) \cong a\,e^{(\lambda-\mu)t} \tag{35}$$

i.e an *exponential increase* with just a smaller prefactor $a$.

This crossing between the decreasing exponential of the bulk of the system and the increasing exponential of an initially tiny minority is one of the recurring motifs in AB kind of systems. This behavior is a direct consequence of the granularity of the A's which is amplified by exponentiation to a much coarser granularity of the B herds / clusters. In turn it leads to macroscopic time cycles with very specific characteristics described in detail in the Figs. 4-8 (summarized in Fig. 9 and illustrated empirically in Fig. 10).

After the exponentials crossing, the system is dominated by the herd / mountain / island generated around the A location. In certain conditions, when the A is not moving too fast the B herd is able to follow the A motion and grow even in conditions where the continuum analysis would predict B's disappearance.

Note that while it is legitimate to say that the emergent B herd follows the A's motion, this cannot be said about the individual B's. Each B has a trajectory which is completely oblivious to the A's. The individual B's don't follow anything! The property of the B islands to "sense", follow and exploit the changes in the A configuration is an emergent property; the individual B's do not have it. In this respect the AB model is the minimalistic (cheapest in the Occam sense) mathematical metaphor to the emergence of living, self-preserving, self-sustaining objects. In the absence of the emergent collective herds, the B population would disappear according to the continuum approximation.

Thus, the B population can survive even in condition where the average/continuity approximations would predict total extinction. In 1 dimension the resilience of the B herd following an A is enhanced by the fact that even if the A is moving very fast, half of the time it will move back to a site that it already visited very recently. Thus a fast A diffusion would still be OK for the survival of the B population because ½ of the time the A will reach a site where the B's have been growing until very recently. Thus the A agent will find there a relatively large B population waiting to restart the growth that was only briefly interrupted (Fig. 8). This would not be automatically true in high dimensions where a fast moving A will mostly go to directions which it did not visit until now and thus are quite empty of B candidates for proliferation.

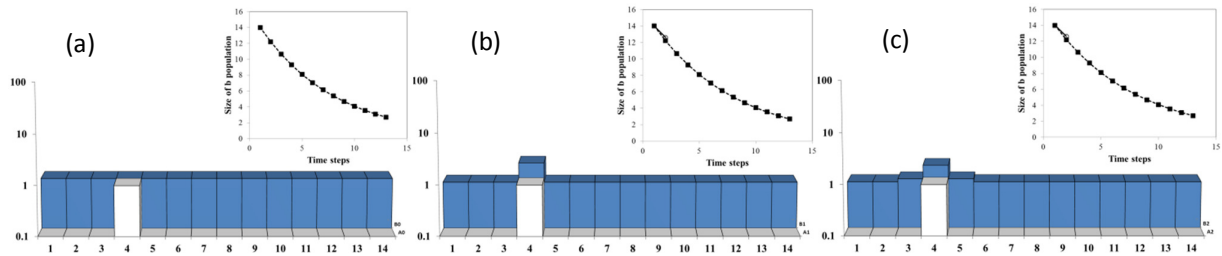

**Figure 4. (a)** describes schematically the time evolution of the B population in the 1 dimensional AB model with 14 sites and 1 single A agent (so the average density of A's is 1/14). The parameters $\lambda = 0.88$ and $\mu = 0.2$ are chosen such that the continuum approximation Eq. 13, $\mathbf{b}(t) = b_0 e^{(a\lambda - \mu)t} = b_0 e^{(0.88/14 - 0.2)t} = b_0 e^{-0.14t}$, is decaying exponentially as shown by the line with filled

square symbol, in the insets at the upper right corner. The position of the A and the distribution and B population are given by the columns in the main figures. The numbers at the bottom label the sites of the lattice. The location of the A agent is represented by a white column. The height of the dark (blue) region at a site represents the amount of B population at that site. **(b)** As the process onsets, the B population at the site where A resides (site 4) increases exponentially as: $b(x,t) = b(x,0)e^{(a(x,t)\lambda - \mu)t} = b(x,0)e^{(1 \cdot 0.88 - 0.2)t} = b(x,0)e^{0.68t}$. At all the other 13 sites the B population decays exponentially as: $b(x,t) = b(x,0)e^{(a(x,t)\lambda - \mu)t} = b(x,0)e^{(0 \cdot 0.88 - 0.2)t} = b(x,0)e^{-0.2t}$. **(c)** The process of diffusion takes place. 10% of each B population is moved to the left and 10% to the right-hand neighbor. In the insets, the size of the B population as result of the discrete model is shown by the line with empty circle symbols (the line advances with each simulation step). In the first 2 steps the continuous estimation (full symbols) and the discrete one (empty symbols) overlap, which shows that the continuum result is a good approximation for the first 2 steps.

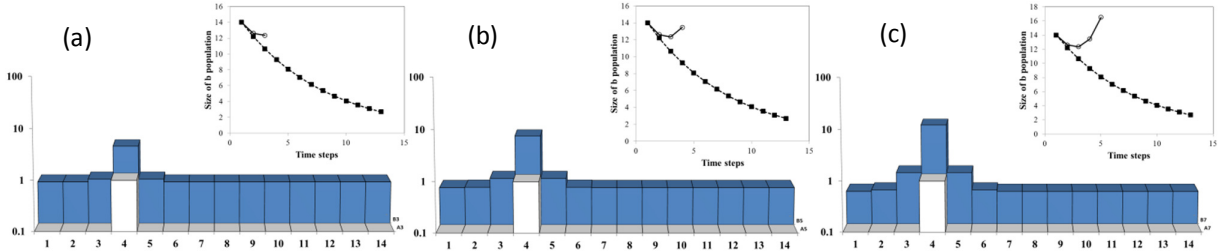

**Figure 5.** As time advances (**(a), (b)**) the continuum approximation (black squares) becomes inadequate. The average of the 14 exponentials $\langle e^{(a(x,t)\lambda - \mu)t} \rangle$ becomes more and more different (larger) from the exponential of the average growth rate $e^{\langle a(x,t)\rangle \lambda - \mu}$. This is seen in the departure of the line with empty symbols representing the true process from the line with black squares (continuum approximation). One can also see the reason for it: the decay of the 13 sites becomes insignificant because their contribution to the average / total B population becomes negligible compared with the B population at the A site (4). An exception are the neighbors of the A location which receive by diffusion part of the B's produced at the A location. In Figure **(c)** the process reaches again (and exceeds) the initial value of the total B population but this time concentrated in the neighborhood of the site where A resides.

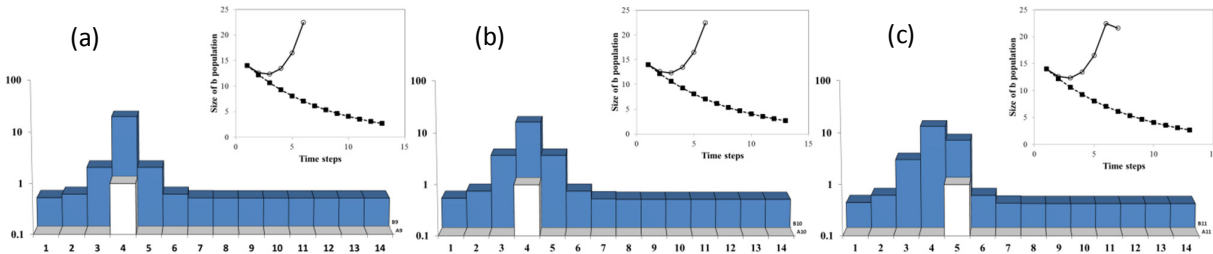

**Figure 6.** **(a)** As long as the A agent maintains its position, the B population at its site (4) increases (it would increase indefinitely if A would stay). **(b)** The diffusion of B generates, increases and widens (at constant speed) the B herd / mountain / island that escorts A. **(c)** The diffusion of A comes into play and the single A jumps on a neighboring site 5. The B population at the old site 4 decays now and the B population on the site 5 grows exponentially.

However, the fast exponential increase in the B population at the new A site 5 is not felt at the beginning because the exponentially increasing part is initially subdominant: $b(x=new,t=jump) \ll b(x=old,t=jump)$. Rather, immediately after the A jump to 5, the fast exponential *decay* of the site 4 that hosted A before the jump is the dominant process, and the total B population decays.

This sharp break between the exponential growth before the A jump and the exponential decay after the jump gives a characteristic cusp shape of the maximum of the total B. We will encounter this characteristic shape in empirical measurements (Fig. 10) as well as in model simulations (e.g. Fig. 2).

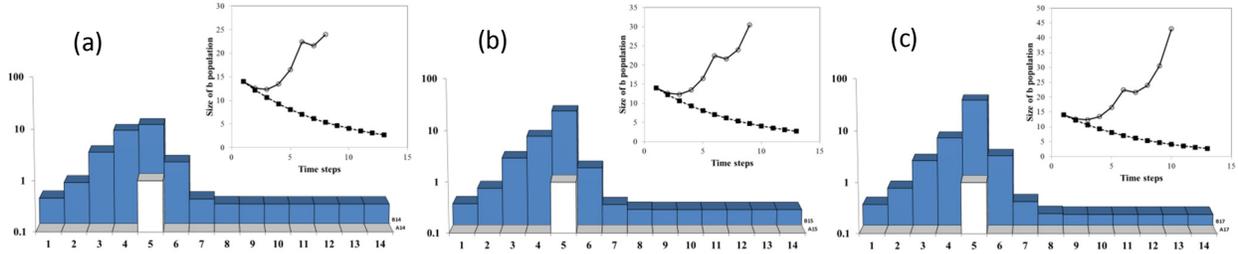

**Figure 7.** **(a)** Immediately after the A jump from site 4 to 5, the system behaved similarly to its behavior at the very beginning of the process: most of the B's were dying without reproducing. However, while the B population at the old site 4 further decreases the B population at the new site 5 increases. Already after 1 step the B level at site 5 equals the level at site 4. This moment is a local minimum of the total B population and corresponds to the crossing of exponentials in Fig. 9. After that moment the total B population starts increasing again.
**(b)** The new site 5 becomes dominant:
$b(x=new,t) = b(x=new,t=jump)e^{(\lambda-\mu)(t-t_{jump})} \gg b(x=old,t=jump)e^{(-\mu)(t-t_{jump})} = b(x=old,t)$ and starts leading the renewed growth of the total B population.
**(c)** The exponential growth sets in and the entire B population starts to be totally dominated by the growth of the site 5 and its overflow to the neighboring sites.

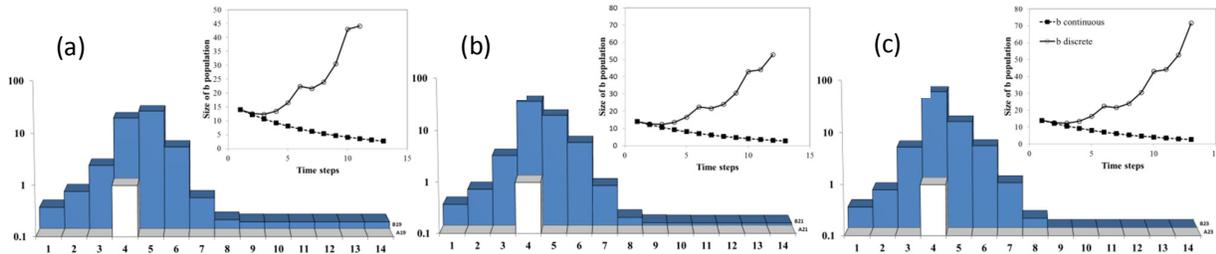

**Figure 8. (a)** A jumps back to site 4 (typically A moves after a time interval of order $\sim 1/D_a$) and B population at the site 4 starts to proliferate. The jump of A creates another cusp in the inset. **(b)** Another crossing takes place: the size of the B population at 4 becomes bigger than the population at 5. This is corresponding to another crossing of exponentials in Fig. 9. **(c)** The system continues to grow due to the exponential growth at the site 4. Due to B diffusion from 4, the sites within a linearly increasing radius join the exponentially growing herd.

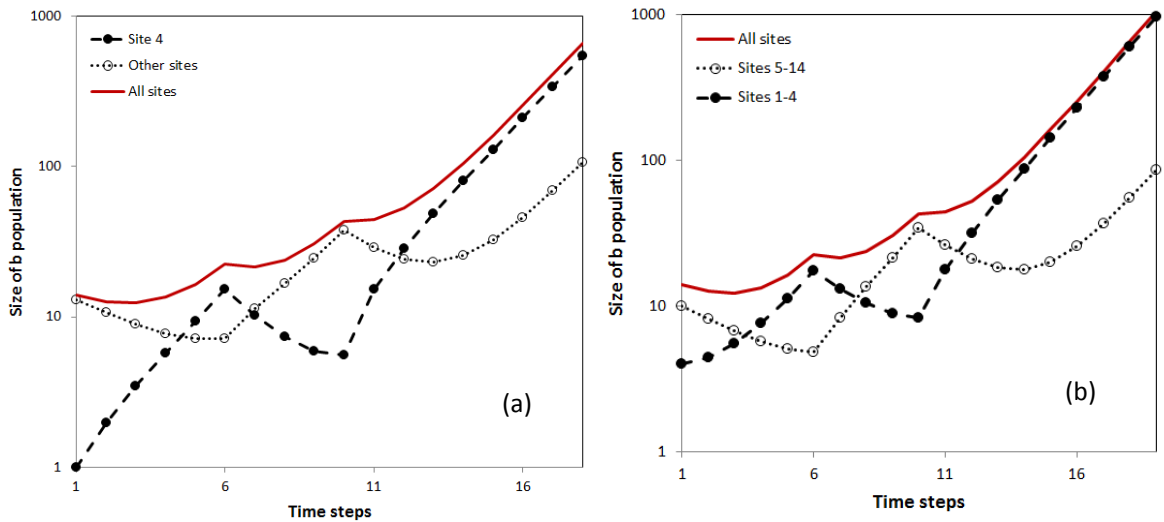

**Figure 9.** The figure details the evolution of the sub-systems during the process Figs. 4-8. The solid (red) line represents the size of the entire B population. **(a)** The dashed line with black circles represents the B population on the site 4, while the dotted line with white circles represents the sum of the population on all other sites. **(b)** The dashed line with black circles represents the b population on the sites 1-4, while the dotted line with white circles represents the sum of the population the sites 5-14. By comparing (a) and (b) it becomes clear that grouping of the sites for the crossing exponentials plot requires only that the sites 4 and 5 are in separate groups (the others sites cannot change the crossing exponentials behavior).

The agent A resided during the time steps 1-5 and 10-18 at the site 4. Between 6 and 9 the growth agent A stayed at the site 5.

Each move of A (t=6, t= 9) produce a shock to the system: the cusps of the solid line.

Immediately after the shock, the growth rates of the dotted and dashed lines diverge.

This is followed by the crossing of the dotted and dashed exponentials (t= 4, t=7, t= 12).

At the crossing time the total B population b(t) reaches its minimum.

Eventually, the growth rates of the dotted and dashed lines equalize (see for instance the times after t=14 in Fig. 9). This can be explained either algebraically through the Frobenius-Perron theorem or geometrically through the diffusion of B's throughout the system (Fig. 11 and the empirical illustration Fig. 12). We call this phenomenon of eventual uniformization of *growth rates* across the entire system "alignment". This is because the term "convergence" refers in economics literature to the convergence of the *GDP per capita* itself which does not happen in our model (nor in reality).

When applied to economic systems (e.g A ~ productivity / skill , B ~ capital) the analysis in Figs. 4-9 shows that as opposed to the usual depiction of the economic cycles which have smooth maxima and minima, the maxima in the AB model are often cusps that separate smooth periods that can be parameterized by just 2 exponentials whose crossing marks the cycles' minima.

Thus while the exact function describing the system time evolution cannot be predicted in detail, the space of functional forms that the growth of economies, populations, technologies can assume is quite limited. In the cases in which the dominant sector is being replaced, one observes a set of cusps that separate crossing exponentials. Each of the exponentials represent the contribution of a specific sub-system: the old decaying one or the new growing one. In the case in which the dominant sector remains the same, the behavior of the system just repeats the variation in that sector. Thus by recognizing the fact that an economy is made out of discrete, granular components allows one to characterize the size and shape of its time fluctuations. This is a tremendous constraint for the space of allowed / expected / predicted evolution functions. In turn, this can be transformed into a powerful forecasting, steering and policy making tool.

So while we do not know when a cusp / shock will take place (when an A will move or when an island disappears) we know that it will be followed by a crossing exponential period. If it is not interrupted by a new shock this phase will in turn be followed by an alignment period in which all parts of the system increase in time exponentially with the same growth rate. The fact that one obtains in Fig. 9 basically the same graph shows that the crossing exponentials shape is robust to the exact partition of the system in "old" and "new" parts. Thus the recognition of the system components limits the number of effective degrees of freedom of a system from a macroscopic one to just a few. This makes its forecast and steering optimally efficient. It is better than both
   - the methods based on an excess of micro-variables and agent and
   - the methods based solely on macro-variables.

The sequence of cusps and smooth crossing exponentials is a hallmark of the AB model (see in Fig. 2 the actual cusps and crossing exponentials episodes in a simulation not limited to a "single A" approximation). This behavior is encountered in many autocatalytic multi-agent systems in many domains. See the realization of cusp-crossing exponentials episodes in the real data of Fig. 10: even small shocks that do not produced significant crises, turn out to be the dominant factors behind the economic fluctuations. For instance in the Fig. 10, one can see that the entire period 1990-2008 can be fitted perfectly by a sequence of 5 shocks / cusps and the 4 crossing-exponential periods that they separated.

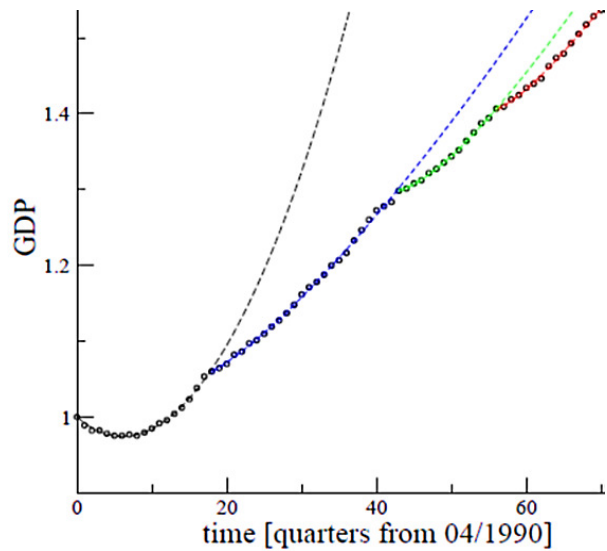

**Figure 10.** Scaled data of real quarterly GDP of Great Britain from 04/1990. From [26]. In a high resolution time series of Britain 1990-2008 the entire range of 70 quarters was fitted with high precision in terms of 4 cross-exponential episodes limited by cusps representing shocks / crises [26]:

- the first at the 1990 recession
- an additional shock in 1995 connected to the real estate bubble burst
- the dot.com bubble burst in 2001
- an additional small one in 2004 and
- the 2008 recession.

(http://creativecommons.org/licenses/by/3.0/).

*5.2 Alignment, predictions and empirical validation*

It has been observed during the numerical simulations and proven using the Frobenius-Perron theorem that if the time between significant changes in the $a(x,t)$ configuration, is large enough, then the B herd / island/ mountain around the dominant $a(x,t)$ region generates and sheds B's that diffuse throughout the entire system.

As a result, for a finite system and after a long enough time, the number of B's grows with the same exponent at all locations throughout the entire system. The pre-factor at each location depends on the flow of B's arriving from the growth center to that particular site. In the case of an economy it depends on the capital flow / payments between the regions, individuals, sectors (in the case of the sectors the flows are related to the elements of the Leontieff matrix). A "single A" computer simulation illustrating such processes is shown in Fig. 11 which displays the main features of the AB dynamics:

- The initial divergence of the growth rates in various parts of the system:
    - around the A location one sees an exponential increase in the B numbers
    - at locations distant from A the B numbers decrease
- One sees "small" ups and downs related with every jump of A on a neighboring site (producing a temporary decay in the B population). Locally those fluctuations are quite significant (factors of 2) but on the background of the macrodynamics of the growing "mountain" they look small on the logarithmic scale.
- As time passes the B herd around A increases both in height as well as in radius. In fact all the points reached by the growing B "mountain" grow with the same exponent and with prefactors decreasing exponentially with the distance from A. Thus the radius increases linearly in time.
- As long as the growth mountain does not reach the size of the entire system, one has 2 distinct regions:
    - the region reached by the "mountain" which increases exponentially.
    - The sites that are far from A and were not reached yet by the growing mountain. They are still decaying.
- When the radius of the B crowd reaches the limits of the system, then all the locations grow with the same growth rate: one has "alignment".

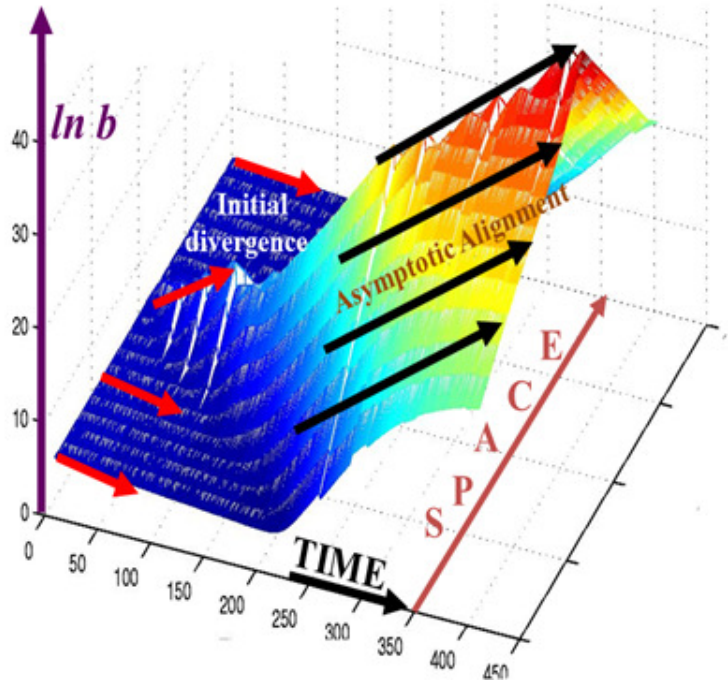

**Figure 11.** The graph depicts a one-dimensional "single A" simulation of the type described in Section 5.1. The axis from left to right denotes the time while the axis going in the depth of the figure represents the 1 dimensional space. The vertical axis measure on the logarithmic scale the number of B agents at each space point $x$ and time $t$: $b(x,t)$. One sees that at the beginning the small region around A is growing while all the rest is decaying. Moreover one sees occasional collapses of the central peak by a factor of 2 each time A changes its site. Eventually the B "mountain" surrounding A expands (linearly in time) until it reaches the confines of the system. At that stage one reaches "alignment" the entire system grows exponentially with the same growth rate (modulo the local temporary fluctuations produced by A jumps between sites). (Source: ref [38])

A real life realization of this scenario is seen in Fig. 12 based on the measurements of the number of enterprises at various locations across Poland:

- before the capitalist autocatalytic dynamics took place the number of enterprises (also GDP and other measurables) was distributed randomly across the country. The spatial correlation in Fig. 12 is seen negligible in 1989.
- after the liberalization the AB kind of dynamics was installed (with A the education years per individual and B the number of enterprises/GDP). The highly educated locations started to develop fast and extend their influence by trade / employment to the geographically neighboring regions. One sees that already by 1992, the radius of their influence reached around 50 km. By 1997 the influence reached the entire country and all locations ended up with the same growth rate. Still in absolute terms the various regions remained separated by large factors in GDP per capita.

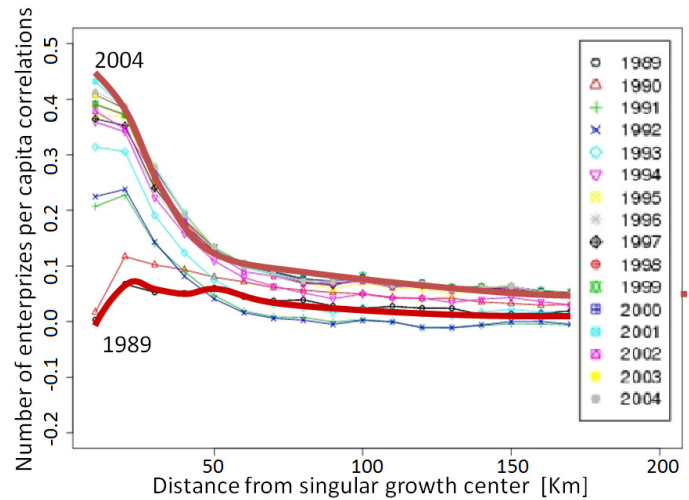

**Figure 12.** The effect of diffusion of growth shown on the Polish economy data (number of enterprises per in different counties located at a specific distance (x-axis) from the location of the singular growth). From: [25]. The graph is exactly reproducing (the right half) of the profile of the growing "mountain" in Fig. 11. The borders of Poland were reached by the growth wave generated in the growth centers already 5 years after liberalization. One should emphasize however that while the growth rates eventually converged, the absolute values for the production / income at different locations remained separated by large factors (up to 2 orders of magnitude) as predicted by the AB model. This might explain the failure of the neo-classical economic measurements to detect any convergence [27].

*5.3 Sectors granularity, crossing and economic cycles [28]*

According to the scenario above, after a significant change in the $a(x,t)$ configuration the dynamics of the system is dominated by just 2 terms in the sum Eq. 13: the old largest $a(x,t)$ before the change and the new largest $a(x,t)$ after the change.
  - At the old location, the dynamics consists of the exponential decay of the dominant B population that used to grow at that location before the change.
  - At the new location, one has an exponential increase of the initially very small B population existing there.

Similarly to Fig. 1, initially the fast growing B population at the new location is too small to affect the systemic dynamics which is still dominated by the (exponential decay of the) large herd at the old location. However, eventually the decaying exponential at the old location and the growing exponential at the new location cross and the total B population makes a turn up. So instead of the spatial uniformity predicted by the differential equations 10-11 the most singular and improbable fluctuation of $a(x,t)$ are the ones dominating the logistic dynamics. We will give below some examples of the real economy where such crossing of leading industrial sectors have been observed.

In such real systems, which are not infinite and thus can receive A's from outside the system (or loose A's to the exterior), fluctuations can appear not only by the A's moving / changing location but also by

A's appearing or disappearing. This might correspond to opening or freezing of foreign markets or to gains or losses of oversea investments. We will see below such examples too.

Fig. 13(a) shows the crossing, and then alignment, of different economy sectors in Portugal, following the carnation revolution of 1974. The crossing happens between the "old" sectors, traditionally encouraged by the austere regime (construction, retail, transport, communication, finance) and the "new" (consumption) sectors (hotels, services, health). After the liberalization, the old sectors started slowing down and the new sectors started taking off. While this was a good thing for the living standards of the population, and eventually the economy recovered, the shock had an immediate negative effect: the GDP followed the old sectors and took a dip in 1974-1977 to recover only later following the growth of the new emerging sectors. One sees also the alignment: the black (empty squares) and red (full circles) lines are parallel 1977-1985.

Fig. 13(b) describes the following crossings and dips in 1985, 1994, 2004. Note that in order to describe the 1985 dip in terms of crossings one has to use a different separation in sectors than the one used for 1974 (the crossing of 1985 just doesn't appear with the sectors partition used in 13(a)).

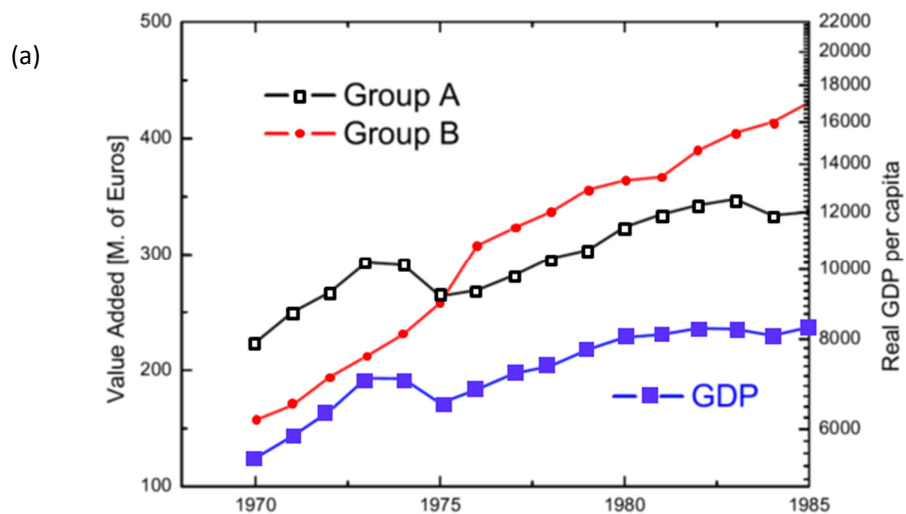

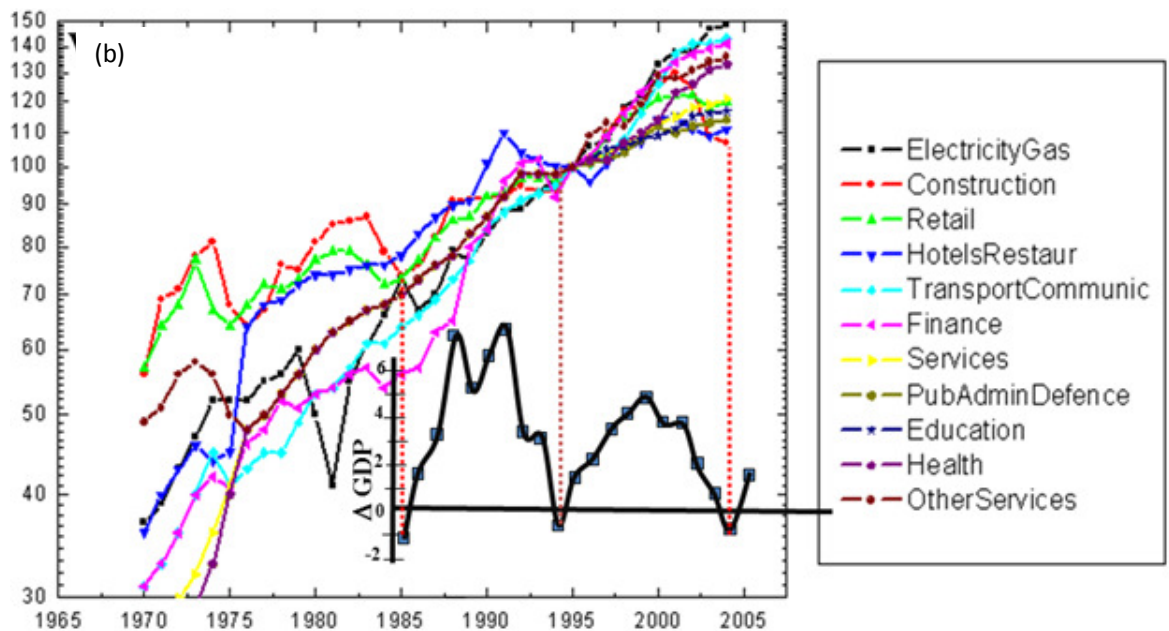

**Figure 13.** Crossing of different economy sectors in Portugal 1970-2004 and corresponding GDP dips. **(a)** documents the aftermath of the carnation revolution.
The black (empty squares) line representing Value Added in manufacturing, construction, transport, finance went down while the red (full circles) line representing VA in education, services, health, hotels, restaurants went up. Their crossing corresponded to a dip in the blue (full squares) GDP line.
**(b)** shows a more detailed picture of the sector's evolution.
The inset, representing the change in GDP is aligned in the time direction to the sectors added values such that one can see the coincidence of the GDP growth minima with the crossing of the dominant sectors:
- the minimum in 1985 corresponds to the construction sector loosing to the Hotels / Restaurants sector (the inverting of the 1975-1976 crossing)
- the minimum in 1994 corresponds to construction, health, energy, Finance and other services dislocating Hotels / restaurants.
- the minimum in 2004 corresponds to collapse in constructions with respect to all and energy taking the lead.
In addition note that between the shocks the growth rates of the different sectors tend to align as expected theoretically from our model.
The main figure is from [28]. Data source of inset IMF WEO http://static.seekingalpha.com/uploads/2009/1/12/saupload_portugal_gdp.jpg.

As explained above, since most of the real systems are open subsets of "the entire universe", it is quite natural to consider the possibility that A's and B's enter and exit it (e.g. capital in and out). As opposed to the abstract AB model, in real empirical systems A's are not necessarily conserved: money / capital can just disappear from a country, or appear there because of foreign/exogenous events. Thus in addition to processes depicted in Figs. 4-9, where the disappearence of an A from site 4 is always accompanied by its appearence on site 5 or 3, there are cases where one just have

(dominant) islands shrinking or expending without any other sectors / herds/ islands playing any role or undergoing any change. With this allowance for A's entering and exiting the system, one does permit fluctuations which have an underlying structure simpler then crossing: the total GDP just follows the variation of the dominant sector (or of synchronized effect of a bunch of sectors). We compare these two cases (crossing vs simple collective object shrinking) in Fig. 14. While the first dip (1980) in the South Korea Economy appeared because of the crossing the old dominant sector (Public services) with the new sector (Manufacturing), the second dip (1998) is a result of a crisis that did not include any sector crossing. The change in GDP reflects mainly the dip in the dominant manufacturing sector. This is even more a demonstration of the role of the components granularity in the time fluctuations of the macroeconomy. It also preserves the principle which becomes crucial in Section 6 that the fluctuations of the collective objects are (stochastically) proportional to their own size. Thus the size of the sectors determine the magnitude of the macroeconomic fluctuations.

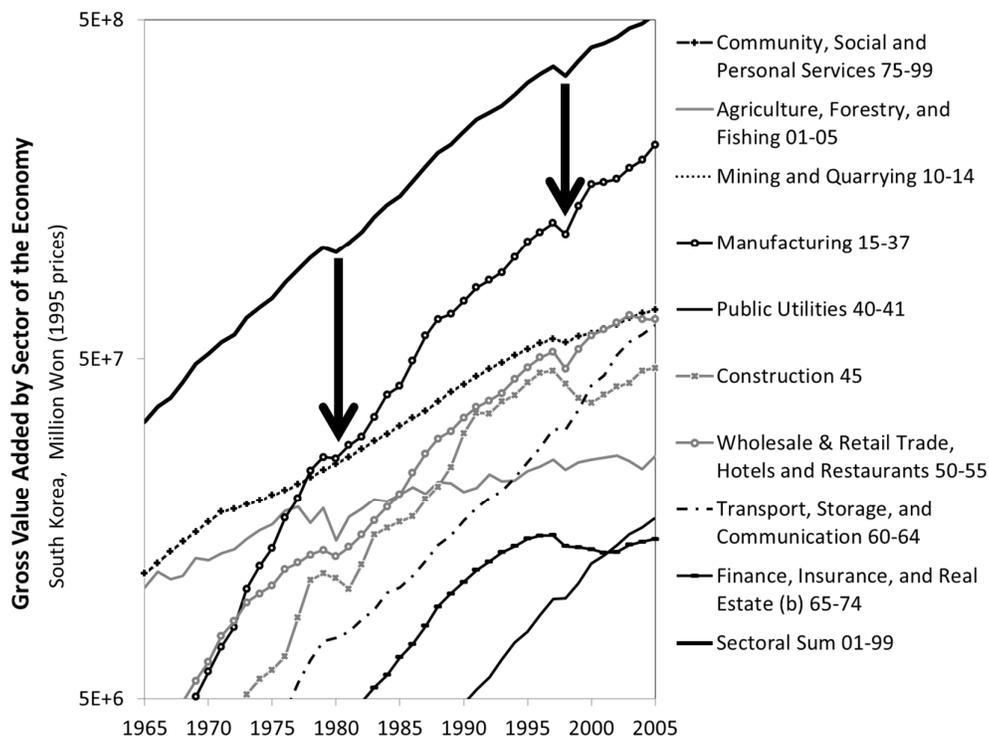

**Figure 14.** Represents the gross added value per different sectors in Korea 1953-2005. The sectoral data are taken from: http://www.rug.nl/research/ggdc/data/10-sector-database. In the 1950's South Korea was one of the poorest countries in the world, while today it is a developed country. The growth of the industrial sector was the main reason for economic development. Since 1980, when the manufacturing sector overtook the previously dominating sector (defence, education and other social services), it stays the dominating sector and most of the other sectors are dependent on its growth. Note that the crossing of the two sectors corresponded to the minima of the total sum of the entire economy (marked by the first arrow from the left).

In 1997-9 Korea was hit by a strong crisis. This exogenous shock had been relatively quickly repared in the dominant Manufacturing sector. Thus even though the sub-dominant financial and real estate sectors suffered much longer, the GDP followed the domiant sector and recovered swiftly.

*5.4 Components' granularity determine time fluctuations [29]; Globalization, inequality and instability*

The fluctuations due to the variation of a single or a couple of sectors as studied in the previous section are illuminating for the kind of effects the AB model offers but do not exhaust them. One may consider the set ("collective") of collective objects that emerge in an AB system and characterize the time fluctuations of the total B population in terms of their size distribution.

In particular one may consider systems with coarser or finer granularity and study their effect on the system's time fluctuations:

-large clusters / herds mean larger amplitude of the time fluctuations of the system.

-small clusters imply system fluctuations close in nature to the Gaussian fluctuations.

In particular one can tune the size of the herds / clusters by varying the competition radius between the B's. For a static A configuration and infinite competition radius one can prove that the only collective object that survives is the B herd around the largest A fluctuation [29]. For diffusive A and finite competition radius the result is milder but the system is still dominated by very few B herds.

When interpreted in terms of economic variables this translates into a connection between globalization, global wealth inequality (capital concentration in a few large fortunes) and economic instability (large – possibly catastrophic - fluctuations). On the other hand, if one tries to eliminate wealth inequality beyond a certain degree, this has negative effects on the GDP. Those effects are displayed in Figs. 15-17:

- In Fig. 15 one compares the B distribution in the presence and absence of long range competition. The same A distribution (Fig. 15(a)) can lead to many small herds (wealth concentrations) (Fig. 15(b)) or to a few large herds (Fig. 15(c)) depending on the radius of competition between the B's.
- In Fig. 16 one sees that extremely large, extremely rare collective objects lead to very large fluctuations in the system time fluctuations.
- In Fig. 17 the amplitude of the fluctuations is connected to efficiency of the A's to produce large B populations. In particular systems with no large collective objects have very smooth, stable dynamics but at a very low B total population. On the contrary systems with very unequal B distribution can generate occasionally a large total B population but the fluctuations are catastrophic.

Thus Fig. 17 illustrates 3 possible regimes:

1. a regime where the competition is local. Large clusters are thus locally suppressed and the clusters are typically small and densely spread throughout the system. In these cases, the granularity of the system components (individual wealths, companies, trade niches) is fine, the growth is muted and the system fluctuations are limited (the lowest, "flat", line in 17(a) and the Latvia example in 17 (c)). One sees that in Latvia, quite a number of years after liberalization, there were not significant fluctuations but it took more than a decade to reach back the pre-liberalization GDP level.

2. A regime where the competition is global. The clusters (individual wealth , capitals) are large and sparse. In these cases the granularity is coarse (the entire wealth is concentrated in the hands of a few oligarchs) and the time fluctuations are wild with periods of very large B population (wealth) and very large collapses. This is seen in the thin (red) line in Fig. 17(a) and the Russia and Ukraine graphs in 17 (b): at times they reached GDP growth rates of more than 10 % a year but also often negative growth rates.

3. An intermediate regime with moderate competition radius and quite good and stable growth rate. The inequality between the clusters is described typically by a normalizable Pareto-Zipf power distribution ($\alpha > 1$). Consequently the time fluctuations of the system are described by a random walk with steps sizes parametrized by a power law with the same exponent $\alpha$ that parametrizes the clusters granularity. Thus the fluctuations are fractal but typically not catastrophic. This is seen in the dashed line in Fig. 17(a) and in the Poland example in 17(b). The fluctuations are moderate and the average B number is high. The growth rate is high but does not reach the peaks of Russia and Ukraine. On the other hand it never gets negative.

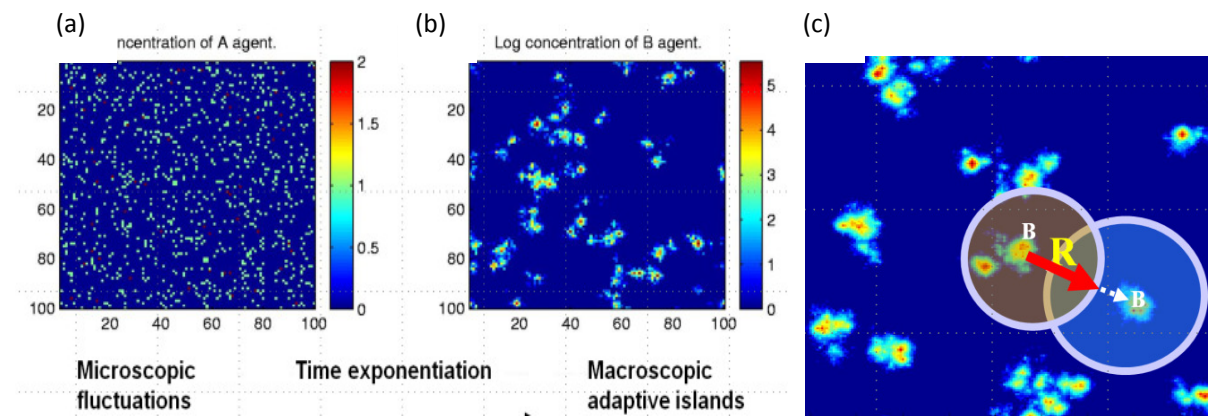

**Figure 15.** Larger competition radius leads to coarser granularity (competition globalization leads to wealth localization). In the absence of competition (or with only local competition) the spatial distribution of herds / islands is solely dictated by the A configuration (Fig. 15(a)) and the herds / islands can have any position (Fig. 15(b)). In this case the sizes of the herds / clusters fulfill typically - in a wide range of parameters - a power law: the number of clusters with B population larger than a size $b$ is proportional to $b^{-\alpha}$ with $\alpha > 1$. However if the B's do suppress one another when they are within a certain radius R one from the other, then the islands inhibit one another typically over the same radius R (Fig. 15(c)). Thus the islands do not get typically closer than R one from the other. If one spreads the competition / suppression of other B's over a larger radius, one obtains herds / clusters

that are rarer but larger. Thus the granularity of the B population is coarser. While in the absence of long range competition the distribution of the island population sizes follows a power law with $\alpha > 1$, for competition over large radiuses the exponent becomes $\alpha < 1$. If one imagines that the clusters are the capital / wealths accumulated by some lucky individuals / firms, then this corresponds to a Pareto wealth distribution which becomes oligarchic in the $\alpha < 1$ range. (Source: ref [38])

Thus the theoretical predictions and empirical observations indicate a clear connection between the time fluctuations in the total B population in the system and the size distribution of the B islands at given time. This corresponds in real systems to connections between the time fluctuations of growth systems and the granularity of their fixed-time density distribution (granularity of the B islands).

So in a very general sense the appearance and disappearance of entire bunches of wealth (belonging to companies, geographical locations or individual persons) dominates the dynamics of the system. Knowing their size distribution allows to predict the time fluctuations of the system: the distribution of the fluctuations' amplitudes will be identical with the size distribution of the wealth clusters composing the system at any given time.

In the present section we discussed large time scales and large amplitude macroeconomic events of the type of economic cycles. In the next section we will see that the connection between time fluctuations and structural granularity governs also the short time fluctuations of the economy as measured for instance by the stock market fractal exponent.

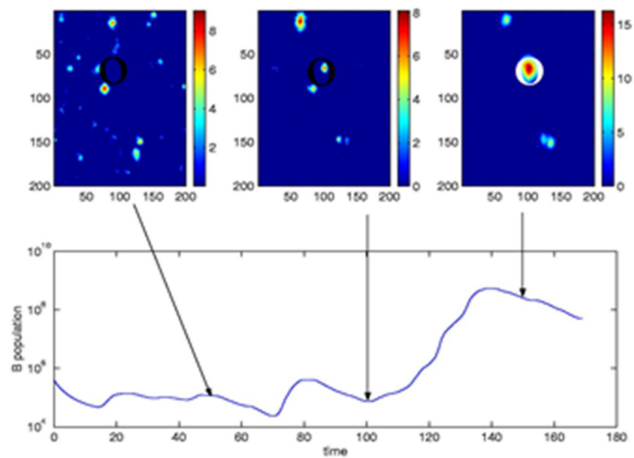

**Figure 16.** According to the Fig. 15, the granularity of the collective objects (individual wealth, firm sizes, sizes of trade niches, local city and regions budgets) depends on the details of the competition radius. In turn, the granularity of the collective objects dictates the time fluctuations of the system: whenever a herd / cluster disappears / grows / shrinks the total B population changes by the same amount. Since in the AB model the growth / shrink of a cluster is typically proportional (via a random number) to its own size (in particular a cluster can disappear completely and it grows / decays exponentially) the total B population is a random walk with steps

sizes distributed by a distribution similar to the clusters' sizes: system granularity determines system time fluctuations intermittency. (Source: ref [38])

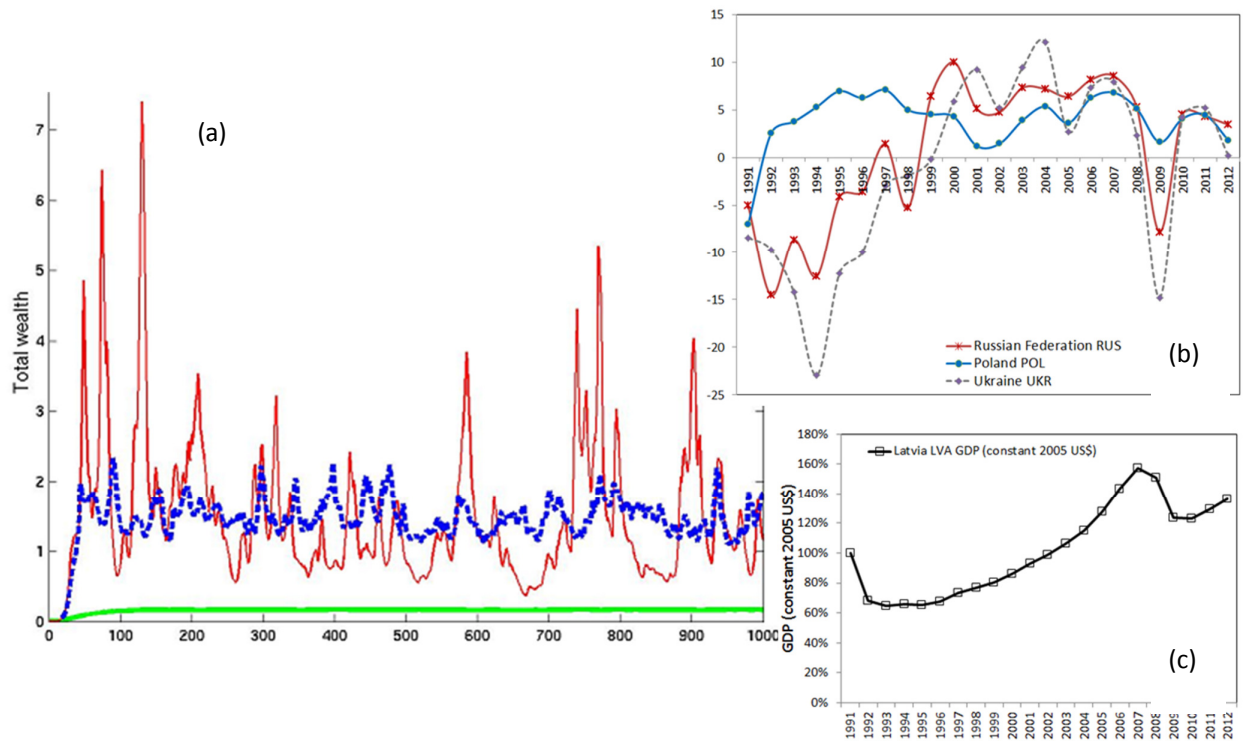

**Figure 17.** According to Fig. 16, the fluctuations in an AB system depend on its granularity. **(a)** shows 3 graphs that characterize systems with different fluctuations and granularity levels. The bottom (thick, flat) line represents a system with very small fluctuations and small granularity. The dashed line represents a system with intermediate fluctuations and granularity. Finally the third (thin) line represents a system with very high fluctuations and a very polarized system (in which very large islands/rich individuals exist). **(b)** shows GDP growth rate for 3 countries: Russia (solid line with asterisk), Ukraine (solid line with circle) and Poland (dashed line with rhomboid)**.** Russia and Ukraine are known for their high granularity. The graph shows that they are also characterized with large the growth rate fluctuations. **(c)** represents GDP changes in Latvia. Although the changes are very large on a large scale, the fluctuations are almost invisible.
http://www.focus-economics.com/en/economy/charts/Latvia/GDP . Data source for countries' GDP: http://data.worldbank.org/country .

## 6. Collective objects fractal exponent determine fractal time fluctuations exponent

One of the main properties of the AB model is the connection between the granularity of the system components (herds, islands, individual wealth, sectors, economic regions) and the system temporal fluctuations. This involves a relatively complex interplay between the various features that the AB system displays:
- emergence of collective objects $i$ that compose the system: herds / mountains / islands etc

$$b(t) = \sum_i b_i(t) \tag{36}$$

- the collective objects have an autocatalytic dynamics: the time variations in their sizes are proportional to their sizes themselves (in particular they can appear and disappear entirely).

$$\Delta b_i(t_i) = \eta_i(t_i) b_i(t_i) \tag{37}$$

where $\eta_i(t)$ are random numbers of order $\lambda$ or $\mu$, depending on the details of the object environment (e.g. the appearance or disappearance of A fluctuations) and $t_i$ is the time at which the change in $b_i$ took place.

- It is well known that a random multiplicative dynamics of the type Eq. 37 implies (in a wide range of conditions) a skew size distribution of the Pareto type [30]:

$$\text{Prob}(b_i(t) > b) = b^{-\alpha} \tag{38}$$

where the Pareto exponent $\alpha$ depends on the details of the diffusion $D_b$ and the discretization of the B particles. The same property is expressed by the Zipf relation when one orders the $i$'s in descending order of their size

$$b_i(t)/b_1(t) = i^{-1/\alpha} \tag{39}$$

- According to 36, 37 and 38, the time variation of the total B population $b(t)$ will be

$$\Delta b(t) = \sum_i \Delta b_i(t_i) = \sum_i \eta_i(t_i) b_i(t_i), \tag{40}$$

i.e. a random walk with individual steps $\eta_i(t_i) b_i(t_i)$ distributed by a Pareto-Zips distribution 38-39.

- The relation 40 relates the time variation of the total B in the system $\Delta b(t)$ (e.g. the wealth / capital invested in the entire system / the stock market index of that particular economy) to the granularity of the collective objects composing it $b_i(t_i)$ (e.g. the individual wealth). More precisely the properties of the random walk that $b(t)$ undergoes depend on the size distribution of its components $b_i$.

  In particular one can find 3 dynamics regions depending on the values of $\alpha$:

1. If $\alpha > 2$ then the fluctuations of $b(t)$ are small and very similar to the Gaussian ones and are not causing significant macroscopic effects. The width of the probability distribution of $b(t)$ over a time interval $t$ grows as

$$\sigma(t) \approx \sqrt{t} \tag{41}$$

which, given the fact that the total integral of the probability has to remain 1, throughout the evolution of the system, implies that the height of the central peak of probability density $P(\Delta b(t) = 0, t)$ [i.e the chance that $b(t)$ returns to the same value after a time interval $t$) decays as:

$$P(\Delta b(t) = 0, t) \approx 1/\sigma(t) \tag{42}$$

i.e

$$P(\Delta b(t)=0,t) \approx t^{-1/2} \tag{43}$$

2. If $\alpha < 1$ the probability distribution (Eq. 38) gives an infinite $b_i(t)$ average. This means that one is in a situation described in the previous sections where a very limited number of sub-systems (sometimes just the dominant one) contain most of the B's and thus control completely the $b(t)$ dynamics. The dynamics becomes intermittent: dominated by the most extreme and rare events. As explained above (e.g. Fig. 10) this is a good description of the economic cycles.

3. If $1 < \alpha < 2$ one is in the Levy flights regime (modulo finite size effects that affect the very distant tails of the distribution [31]; there the decay exponent is $2\alpha$ instead of $\alpha$ because in order to have a very large jump in the index in the stock market one needs a very large transaction and thus one needs that both traders are large enough).

 The width of the fluctuations $\sigma(t)$ over a time interval $t$ is dominated by the size of the largest step expected to happen during a sequence of $t$ steps. I.e. the step of a size that has roughly probability $1/t$ to take place:

$$1/t = \text{Prob}(b_i(t) > \sigma) \approx \sigma^{-\alpha} \tag{44}$$

Thus the width one expects the process to achieve after $t$ steps is:

$$\sigma(t) \approx t^{1/\alpha} \tag{45}$$

and because of Eq. 35:

$$P(\Delta b(t)=0,t) \approx 1/\sigma(t) \approx t^{-1/\alpha}. \tag{46}$$

One thus predicts that in a given economy the exponent of the Pareto wealth distribution as extracted from Eq. 39:

$$\alpha = \ln i / \ln(b_1/b_i) \tag{47}$$

equals the exponent in Eq. 46 characterizing the fractal fluctuations of the index of stock market (as a proxi of the fluctuations of total wealth / invested capital) in the same economy:

$$\alpha = \ln t / \ln[P(\Delta b(t)=0,t)/P(\Delta b(t)=0,t=1)] \tag{48}$$

This relation that connects the granularity of the B clusters / herds / collective objects (Eq. 39) (sub-figure 18(a)) to the time fluctuations of total quantity of B's in the entire system $b(t)$ (Eq. 46) (sub-figure 18(b)) is quite precisely verified by the empirical data in 3 different countries: sub-figure 18(c).

More precisely, if one considers that the actual wealth worth of the entire economy is measured at the stock market by the market index, one may try to connect its fluctuations to the distribution of wealth at a given time between the various participants. In particular one would compare the fractal exponent characterizing the market index in an economic system to the Pareto exponent characterizing the distribution of wealth within the system.

Such a comparison is really quite daring as it connects the very volatile dynamics of the stock market to the relatively static distribution of wealth between the individual economic and financial players. Yet, when this confrontation was performed by empirical measurements, the theoretical prediction connecting the exponent of wealth granularity to the exponent of index fractality, has been confirmed very precisely [29] (Fig. 18). Measurements with similar conclusions were performed in [32].

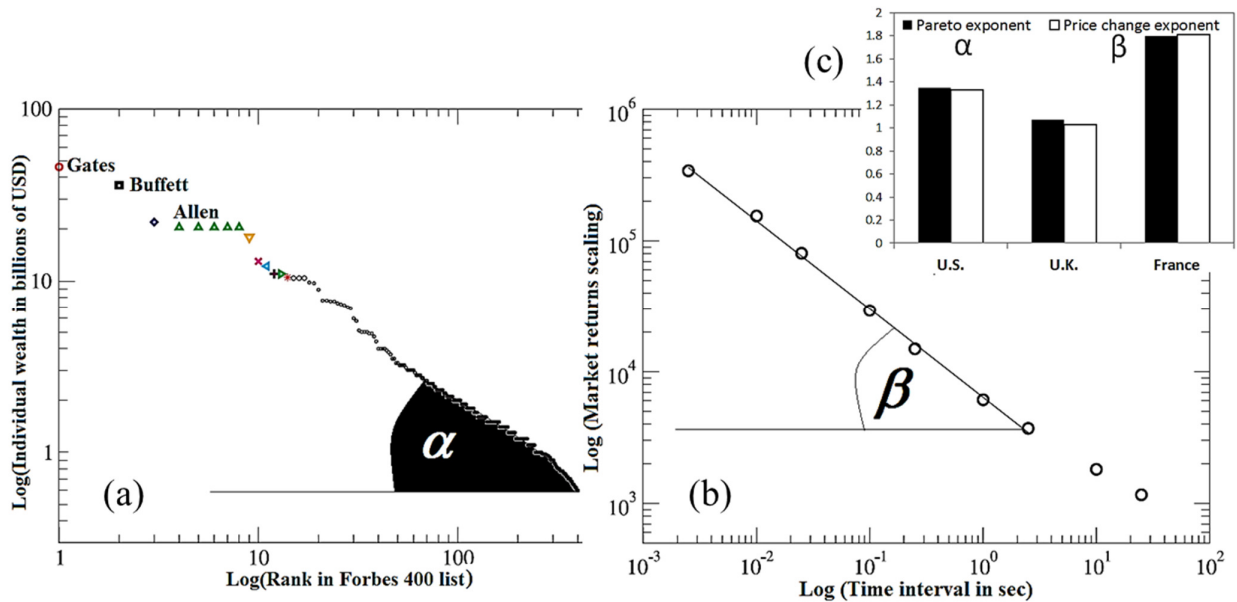

**Figure 18.** Granular structure of individual wealth / capital distribution determines the size distribution of the financial system time fluctuations. **(a)** represents the granular structure of the wealth distribution in terms of the individual wealth of the wealthiest people in US (Forbes 400). It is found that it fulfills a Pareto-Zipf power law distribution $B(i) \approx i^{-1/\alpha}$ where $i$ is the rank of the individual in the Forbes list and $B(i)$ is their wealth. **(b)** represents the fractal scaling properties of the returns $\ln[B(t_0 + t)/B(t_0)]$ distribution $P(\ln[B(t_0 + t)/B(t_0)]; t) \sim t^{-1/\beta}$ in terms of the scaling of the height of its peak $P(B(t_0 + t) - B(t_0) = 0; t) \sim t^{-1/\beta}$. Because the integral of the probability is 1, this also measures the dependence of the width of this probability distribution on the time interval: $\sigma \sim t^{-1/\beta}$. (note that this does not imply necessarily that the very long tail of then distribution decays by the same exponent [33]). **(c)** compares the empirical values of the 2 exponents $\alpha$ and $\beta$ in 3 different countries: One finds quite precisely that $\alpha = \beta$ thereby validating empirically the theoretical predictions of the AB model (or any model (e.g. [31]) with the properties 36-41 ).

## 7. Implications on the emergence and survival of life itself

*7.1 Early search for hints on the nature of the laws underlying the emergence of life*

The agent based / microscopic representation of autocatalytic systems of the type discussed in detail in the previous section provide a deeper understanding of the way the natural selection phenomena act and of their real properties. Instead of the convergence towards an eventless asymptotic state where all species reach maximal fitness, one produces theoretical predictions in better agreement with the empirical observations.

The AB model provides a mathematical metaphor in which a set of chemicals A and B which respect simple rules, rather than approaching a disordered uniform phase, self-organize spontaneously and generate adaptive complex collective objects. Those objects act in a way that dramatically increase their own resilience and the survival of the B populations. The model demonstrates how chemistry can create life in spite of apparent obstructions from the second law of thermodynamics (the law of increasing entropy). Essentially, the AB model is taking a bunch of molecules A and B which naively should reach some kind of thermal equilibrium and become uniform in space and time. Instead of it, they self-organize in collective objects, which clearly do not reach uniform state or even steady state. Those collectives, come alive, die, transfer life to other regions, and most of all act in a way in which they insure their own sustainability by discovering, following and exploiting the random spatio-temporal fluctuations intrinsic to the granular character of the A background.

This scenario answers and overrides Erwin Schrödinger's dilemma expressed in "What Is Life?" that the physics of his time was not sufficient to explain the living organisms. He finally concluded that organisms were repositories of what he called *new physics* [7]. It is now likely that Phil Anderson's "More is Different" [20] is this new physics: no new fundamental laws but new emergence mechanisms in great measure independent and more general than the particular physical and chemical basic laws through which they act in one instance or another.

Yet already half a century ago, before the discovery of DNA, there was a feeling that autocatalytic loops may be the key:

*"The difference in structure is of the same kind as that between an ordinary wallpaper in which the same pattern is repeated again and again in a regular periodicity and a masterpiece [……] which shows no dull repetition, but an elaborate, coherent, meaningful design"* – What is Life , Erwin Schrödinger [34].

The paradoxical immateriality of the laws of emergence which are not in addition yet not included in the fundamental physical laws has been sensed also by Thomas Mann (1875-1955), in the Nobel prized "Magic Mountain" [35].

*"What was life? No one knew. It was undoubtedly aware of itself, so soon as it was life; but it did not know what it was…it was not matter and it was not spirit, but something between the two, a phenomenon conveyed by matter, like the rainbow on the waterfall, and like the flame. Yet why not material? – it was sentient to the point of desire and disgust, the shamelessness of matter become sensible of itself, the incontinent form of being."*

*7.2 The emergence of life teleology from physical causality*

Causality is defined in Wikipedia [36] as "*the relation between an event (the cause) and a second event (the effect), where the second event is understood as a consequence of the first*". It explains an event in terms of a past event – its cause.

Teleology is defined as *"The use of ultimate purpose or design as a means of explaining phenomena."* It explains an event or state in terms of a future state or event.

When explaining the behavior of atoms, molecules one uses *causal* arguments.

When explaining the actions of living creatures one uses teleological arguments: the rabbit will run away from a wolf because the wolf may kill him and eat him up; the amoeba is moving in this particular direction to eat something up.

Teleology is the absolute inverse of what science is supposed to do: to explain in a causal way what is going on. Thus, the jump between the causal explanation which is characteristic for chemistry and physics and the teleological explanation which are absolutely unavoidable in the biological and psychological sciences, has been focus of many studies, starting from the hard sciences and ending with philosophy.

Essentially, very same atoms seem to behave very differently in chemistry and in biology. In the case of a stone they are following explicitly the laws of mechanics while inside of human body we have a behavior which, at least superficially seems the go opposite of mechanics. Even at the most trivial level: when we stand up, we are essentially in a non-stable equilibrium. The trials to follow the line of argument that after all the way the atoms behave in the muscles is perfectly causal has been analyzed in detail in [7] and showed to lead to an infinite causal chain of implications that does not stop or close in a satisfactory form without being short-circuited by a teleological stage.

The relation between causality - the arrow of time - and the law of entropy has been long recognized [37]: "The Second Law of Thermodynamics, discovered in the 19th century, helps define an arrow of time"

Where does the boundary between the causal domain and the teleological domain lie? How does one cross it? The above discussed AB model is a zoom-in on this frontier. It is the frontier between the B-atoms and the B adaptive collective objects. A naïve description of what is going on in the AB model is that the B's are surviving, as opposed to the differential equations prediction, because they are following the A's. However this is not exactly correct: the individual B agents are not following anybody, they have by definition an equal probability of moving in any direction. And that is what they do independently of the positions of the neighboring or more distant A's. The ones that are following the A's, are the herds of B.

According to the naïve estimations, the individual B's should die if $\lambda a < \mu$, because there are strictly speaking not enough A's to generate enough births to compensate for the B's death rate. The way one perceives the survival of B's is in terms of the herds / islands that behave as if they are looking for and exploiting the temporary local fluctuations generated by the A's random diffusion. Thus we have in the AB mechanism a simple, solvable example in which a system crosses the frontier between blindly following elementary rules and adaptively following self-preserving and self-serving strategies. Let us discuss several aspects of it.

**The directed percolation view**

By evoking directed percolation [38] the narrative above can be translated in a precise mathematical language. Imagine the large $a(x,t)$ locations where life can happen are ordered in time Fig. 3. If they are dense enough, the spatio-temporal regions which can sustain B populations can be connected (can overlap) and the B population would be able to continue indefinitely. If the large $a(x,t)$ are not dense enough or the B islands that they can sustain are not large enough, B-life will go extinct.

In field theory terms, this is expressed as a transition towards the directed percolation. Not only the life of individual B's becomes irrelevant to the B population survival but even the perishability of the individual B-herds is not enough to prevent the B population immortality. A possible analogy would be that not only the self-organization of cells in multi-cellular animals insures life perpetuation but also one can afford the death of the individual organisms as long as even one cell is left as a seed for continuing unbroken the chain of reproduction. By the mechanisms of proliferation, life succeeded not only to defeat entropy for 70 years or so and then die, but actually life finds an iterative way for defeating entropy forever. In this context, the metaphor of the civilizations that we gave in Section 1 is congenial to the analysis in [38].

**The cells as a life's instrument to survive**

In this and the following paragraphs we will mention some other facets interpretations of the AB survival scenario.

Imagine a Primordial Ocean where a lot of complex molecules of the protein type (analog to our A's) are created and destroyed. Suppose that suddenly some molecules appear (B, or RNA) which, if given as prime material, a renewable source of proteins, would reproduce themselves . The AB model analysis says that if this Primordial Ocean is large enough, the RNA population will survive, even if the RNA molecule is not very stable, and even if its capability of reproduction is not very fast, and even if the average density of proteins is very low. The AB model even explains how it will survive: they will form adaptive islands which will search, follow and exploit opportunistically the regions of fortuitously high A fluctuations. In other words, one of the mechanisms for life emergence and growth is the formation of spatio-temporal localized proto-cells. The localization of RNA in the places where there are proteins, and after that, the synthesis of the new proteins that can sustain the RNA production amounted to the "invention" of cells. The cell was therefore one of the tricks that life had to invent in order to emerge and survive in a hostile environment which, on average, was very unfit for life and reproduction.

**Species as life's instrument to survive**

One can look at the AB model from another point of view which reveals the "invention of species" as another "trick" of life to defeat the odds of an overwhelming hostile world. Suppose now (very similarly to Sect. 5) that the space where the A's and B's move is not the physical space but the genomic space represented as a network: each node represents a certain genome with a certain RNA or DNA chain and neighboring nodes represent chains that can be obtained from one another by a single mutation. Each B agent on a node represents an agent with a genome corresponding to that node. B's jumping between neighboring nodes corresponds to mutations. Then the A's presence on a node is a way of representing a randomly changing fitness landscape: the nodes with more A's are the niches advantaged at that particular time. They insure that the B's having the genome corresponding to that node multiply faster. In this setting, the story that the AB model tells is that even if the average fitness of the landscape is well below the one necessary to sustain B survival, the B population will still survive in spatio-temporal islands that follow and exploit the local changes in the fitness landscape. The interpretation of this mechanism is that the emergence of species (connected, separated sets of sustainable genomes connected one to another by simple mutations) is the mechanism by which living agents populations survive even in conditions in which the average fitness would imply death. Thus even before the appearance of sexual proliferation that created additional obstacles between the species, life invented them as a "trick" to insure survival in an, otherways hostile and unpredictable, fitness space.

**Natural Selection of the Medium Fitness as Life's instrument to survive**

A third survival mechanism that the AB model implies is selection of the **medium** by the animals. The commonly held idea, starting with Darwin is that the medium is selecting the species. This is a valid but very slow mechanism for survival. It takes at least 10 generations, because the medium can only act on the rare advantageous mutations after they happened. Opposed to this slow mechanism, there is a much more immediate mechanism: The selection of the optimal medium by the animals. For this mechanism to hold, it is not necessary for the animals to choose consciously or preferentially the better locations. This happens automatically even for the most passive animals. The only property they need to have is self-reproduction. Suppose there are in the medium locations which are favorable for the particular species we are discussing and places which are hostile. The AB model brings our attention that the new animals will always be born in the favorable places. Indeed, the mere fact that in the places which are not habitable there will be no animals, means that there will be no parents to sire the new animals at those locations. On the contrary at the favorable locations a lot of new animals will be born since their parents have a large probability to be there. This mechanism may look at the first sight as a tautology but without the increase increased probability of the parent B's to exist in the immediate neighborhood of A, the chances for producing offspring would correspond to the "bare" $\lambda$ rates which are insufficient to insure population survival. It is the "anomalous" local correlations between the A density and B density which bring the "effective" / "renormalized" birth rates to values that insure survival at the large scales (cf Section 4.1).

## 8. Conclusions

Historically speaking, it was only after the AB model properties were discovered and verified, that it became obvious that, when written in the format of the reaction diffusion equations, it reduces to the logistic equation. This was a lucky situation. If it was known from the beginning that the AB model reduces in the continuous approximation to the logistic equation, it would probably not have been studied by means of microscopic simulation. Indeed the previous 200 years old studies of the logistic equation by means of differential equations did not give any indication that it might lead to such far reaching effects.

The AB model emerged as a quantitative, mathematical counterpart to ideas that were expressed in the past by thinkers in various disciplines. The novelty which nowadays research brings is not so much in enhancing the conceptual depth of the earlier authors such as Hegel, Schumpeter or Darwin but in integrating and exploiting the scientific and technical tools developed in the meantime:

- theoretical developments in the complex systems which make the conceptual points much stronger, quantitative and explicit: statistical mechanics, phase transitions, renormalization group, scaling theory, stochastic processes.
- big data availability that allow the massive detailed acquisition, documentation and analysis of empirical systems: microbiologic, ecologic, economic and social.

We hope that the present sample of the generic properties of the autocatalytic, discrete spatially extended stochastic systems might elicit further studies into the mechanisms that ensure the emergence of collective, adaptive, complex objects in systems that otherwise seem doomed to uniform, monotonic uneventful decay.

**Acknowledgments**

We thank our many coworkers that initiated, conceived and developed many of the ideas and results presented in this essay.

**Conflicts of Interest**

The authors declare no conflict of interest.